\documentclass[lineno,3p]{elsarticle}

\usepackage{framed,multirow}

\usepackage{amssymb}
\usepackage{latexsym}
\usepackage{amsmath}
\usepackage{rotating}
\usepackage{graphicx,color}
\usepackage{bm}
\usepackage{color}
\usepackage{multirow}
\usepackage{algorithm}
\usepackage{algpseudocode}
\usepackage{enumitem}
\usepackage{breakcites}		
\usepackage{natbib}
\usepackage{lscape}
\usepackage{array}
\usepackage{longtable}
\usepackage{epstopdf}
\usepackage{morefloats}
\usepackage{algpseudocode}
\usepackage{subcaption}
\usepackage{url}
\usepackage{xcolor}
\usepackage{booktabs}
\usepackage{cancel}

\usepackage{hyperref}
\hypersetup{
    colorlinks=true,       
    linkcolor=blue,        
    citecolor=blue,        
    filecolor=magenta,     
    urlcolor=cyan          
}

\usepackage{soul}

\usepackage{blindtext}

\begin{document}

\begin{frontmatter}
	\title{Three-dimensional multi-physics simulation and sensitivity analysis of cyclic hydrogen storage in salt caverns}%
	
	\author[TUDelftCiTG]{Hermínio T. Honório\corref{cor1}}
    \author[TUDelftCiTG]{Hadi Hajibeygi}

	\cortext[cor1]{Corresponding author}
	\fntext[fn1]{H.TasinafoHonorio@tudelft.nl}

	\address[TUDelftCiTG]{Faculty of Civil Engineering and Geosciences, Delft University of Technology, Stevinweg 1, 2628CN, Delft, The Netherlands}
 
	\begin{abstract}
            Large-scale storage technologies are crucial to balance consumption and intermittent production of renewable energy systems. One of these technologies can be developed by converting the excess energy into compressed air or hydrogen, i.e., compressed gas, and store it in underground solution-mined salt caverns. Salt caverns are proven seals towards compressed air and hydrogen. However, several challenges including fast injection/production cycles and operating them in a system of caverns fields are yet to be resolved in order to allow for their safe scaled-up utilization for energy storage. The present study investigates time-dependent mechanical behaviors of salt caverns, individually and in multi-cavern systems. The stress conditions and model constructions are chosen to be relevant for underground energy storage purpose, and reflect the essential microscopic deformation mechanisms for reliable predictions. Built on these, the impact of model calibration and parameter estimation on cavern-scale performance are investigated. Moreover, the impact of the transient and reverse creep, cavern complex geometry, and non-salt interlayers on salt cavern performance are studied in detail. Importantly, the presented studies include quantification of the interaction of the caverns in a system of caverns. These are achieved by developing an open-source three-dimensional finite element simulator, named as ``SafeInCave'', which also incorporates a comprehensive salt rock constitutive model. The findings provide insights into improving the reliability of numerical simulations for safe and efficient operation of salt caverns in energy storage applications.
	\end{abstract}
	
	
\end{frontmatter}

\section{Introduction}
\label{sec:introduction}
\begin{figure}[!h]
	\centering
	\includegraphics[scale=0.4]{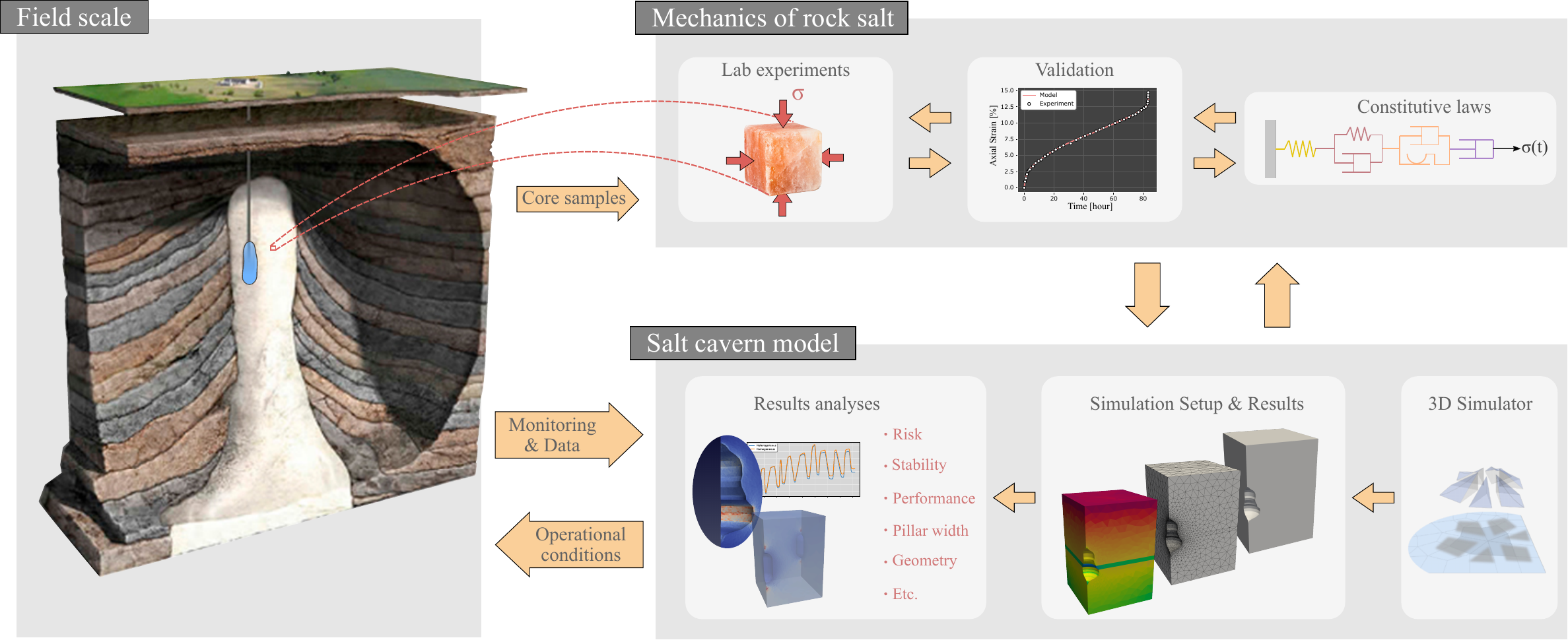}
	\caption{Schematic overview of the modeling and simulation workflow for salt cavern systems.}
    \label{fig:fig_1}
\end{figure}

One of the main challenges in decarbonizing the energy mix is to develop large-scale storage technologies that allow for balancing the consumption and the intermittent production in a seasonal time scale \cite{maia2016experimental}. For instance, 17\% of the wind energy produced in China in 2017 was wasted due to the production-consumption imbalance \cite{fan2019discontinuous}. One of the solutions is to convert the excess energy either in the form of compressed air (mechanical energy) or hydrogen (chemical energy), and then store these gases in underground reservoirs. Although depleted gas reservoirs hold tremendous potential for energy storage \cite{kumar2023comprehensive}, salt caverns are recognized as the most viable option for several reasons. First of all, salt caverns are proven to be applicable, being in use since the 1940s \cite{bays1963use}, for storage of various fluids such as crude oil and its derivatives \cite{lux2009design,koenig1994preparing}, natural gas \cite{tarkowski2021storage}, compressed air \cite{guo2016comparison}, hydrogen \cite{tarkowski2019underground}, radioactive fluids \cite{tounsi2022numerical}, and carbon dioxide \cite{dusseault2004sequestration}. The extremely low porosity and permeability of salt rocks \cite{liu2024optimization} make salt caverns an almost perfect seal, especially for hydrogen storage. In addition, creep and self-healing are favorable rock salt properties that promote cavern stability. However, fast injection/production cycles imposed on salt caverns and the necessity of scaling up these storage systems need to be carefully investigated to ensure safe and efficient operations. For this purpose, lab-scale experiments, development and calibration of constitutive models, cavern-scale simulations, and field monitoring must be integrated into a well-defined workflow, as depicted in Fig. \ref{fig:fig_1}.

Understanding the time-dependent mechanical behavior of salt rocks under relevant stress conditions is the first step to developing constitutive models that can be later integrated into salt cavern simulators. Generally, a salt rock sample shows transient (primary) and steady-state (secondary) creep in response to an imposed constant loading condition. When the sample is unloaded, reverse (transient) creep is also observed. If the stress state lies inside the dilatancy region, long-term brittle failure may occur due to the unstable creation of micro-cracks (tertiary creep), and both permeability and volume are increased \cite{hunsche1999rock,schulze2001development,peach1991influence,peach1996influence}. Conversely, micro-cracks can be suppressed if the salt rock operates in the compressibility region in a process called self-healing \cite{chan1998recovery,zeng2024self}. These phenomena are manifestations of different microscopic mechanisms taking place within the salt crystal lattice, such as dislocation climb, dislocation glide, cross-slip, and solution precipitation \cite{munson1979preliminary,urai1987deformation,urai1986weakening,urai2017effect}. Transient creep, for instance, is caused by a temporary imbalance between the creation and recovery of dislocation movement, whereas steady-state creep is observed when these processes balance each other \cite{cristescu1998time}. These time-dependent microscopic processes are influenced by both stress and temperature conditions. Moreover, the loading rate and cyclic loading (fatigue) are also shown to affect salt rock mechanical behavior \cite{liang2011effect,cristescu1994visco,zhao2022rock,lyu2021study,wang2021experimental}.

The complexities associated with salt rock mechanics render the constitutive modeling development a challenging task. Ideally, constitutive models should be developed based on a thorough understanding of the microscopic processes as it provides more confidence when extrapolating results outside the range of experimental conditions. This is the case for pressure solution \cite{urai1986weakening,spiers1990experimental} and dislocation creep, for which well-established theoretical constitutive models are available and are shown to provide reliable predictions. However, the remaining deformation mechanisms -- transient (and reverse) creep, tertiary creep, dilation, self-healing, etc -- are much more complex to understand at a microscopic level, and a phenomenological (or empirical) approach is often adopted. Besides these complexities, these deformation mechanisms are not all equally important, which is why most authors avoid including all phenomena at once. The widely used LUBBY2 model, for instance, describes only transient and steady-state creep by considering Burger's model with stress-dependent viscosities for the dashpots \cite{heusermann2003nonlinear}. The Hou/Lux-ODS model was developed to overcome some of LUBBY2's limitations related to hardening and recovery, and Hou/Lux-MDS includes damage and healing processes \cite{hou2003mechanical}. The Munson-Dawson (M-D) constitutive model was originally developed for the WIPP salt, and it describes transient creep and three different mechanisms for steady-state creep \cite{reedlunn2016reinvestigation}. Later modifications of this model include reverse creep \cite{munson1993extension}, Hosford equivalent stress \cite{reedlunn2018enhancements}, damage and healing \cite{chan2001permeability}. The transient creep formulation of the M-D model was also combined with the double-mechanisms creep law \cite{maia2005triaxial}, originating the so-called enhanced double-mechanism law using transient function (EDMT model) \cite{firme2018enhanced}. In the Composite Dilatancy Model (CDM), besides transient and steady-state creep, dilatant behavior of salt rock is described, including the effect of humidity on creep \cite{hampel2012cdm}. In \cite{khaledi2016stability}, the Norton creep law was effectively modified to capture volumetric changes in the dilatancy region, and transient creep was modeled using Perzyna's formulation with Desai's yield surface \cite{desai1987constitutive}. The list of models is too extensive, but some important models still deserve to be mentioned \cite{aubertin1999rate,cristescu1993general,hunsche1994kriechverhalten}. Also worth-mentioning are models including damage \cite{deng2020viscoelastic,ma2013new} and fatigue effects \cite{zhao2023creep} in salt rocks.

The number of material parameters within a constitutive model increases naturally as more deformation mechanisms are included. For instance, the viscoplastic model of Desai \cite{desai1987constitutive}, the Hou/Lux-ODS and Hou/Lux-MDS \cite{hou2003mechanical} depend on 11, 11 and 18 material parameters, respectively. As another example, in \cite{reedlunn2018enhancements} it is stated that it is not clear how some of the parameters in the initial M-D model have been calibrated, thus new calibrations were performed. This represents a serious challenge for model calibration and many laboratory experiments are required. For this reason, some authors decide to keep well-established (calibrated) models and add new features (e.g. transient creep) as necessary \cite{firme2018enhanced}. In \cite{tasinafo4782265multi}, a general calibration strategy is proposed based on an optimization framework in which new experiments can be added as they are made available. Although the impact of model calibration against laboratory experiments can be readily assessed, studies on its impacts on salt cavern simulations are not reported in the literature. 

Salt cavern simulations, like any other numerical analysis, are designed to investigate particular problems associated with the different stages of the cavern life-cycle (i.e., construction, operation and abandonment). The cavern leaching, debrining, and operational phases were simulated in \cite{khaledi2016stability}, where different stability criteria were also analyzed. Cavern volumetric closure is usually analyzed during the cavern operation period \cite{reedlunn2016reinvestigation,reedlunn2018enhancements,sobolik2021effect,liu2024optimization}. The subsidence is also a concern during cavern operation and especially after abandonment \cite{buzogany2022development,sobolik2021effect}. Due to salt creep, borehole and casing stability is also a concern \cite{firme2016assessment,orlic2016numerical}. Other studies focus on cavern stability and risk assessment in bedded salt formations, system of caverns and proximity of fractures \cite{liu2024optimization,ma2013new,wang2015allowable,maia2019potential,peng2022comprehensive,sobolik2021effect,xing2015horizontal}. Cyclic operations, which are particularly relevant for renewable energy storage, have been investigated in many studies \cite{coarita2023hydromechanical,zhao2023creep,wang2022long,khaledi2016analysis}. However, few parametric studies have been conducted to investigate, for example, the importance of pressure solution creep \cite{kumar2022influence,sobolik2021effect}, the impact of interlayers and cavern complex geometries \cite{ramesh2021geomechanical}.

Designing relevant simulations and ensuring reliable results are the main challenges faced in numerical analyses. The quality of salt cavern simulations can be measured by the distance between the simulation results and field measurements (i.e., reality). This distance is affected by many variables, such as the choice of appropriate constitutive models and numerical schemes, appropriate lithological characterization, cavern geometries, boundary conditions, etc. Understanding how these variables affect the simulation results is essential to close the distance between simulation and reality, thus building trust in numerical simulations. If this distance is interpreted as a function to be minimized, sensitivity analysis allows us to identify which direction to move in order to improve results. From this perspective, sensitivity analysis can be compared to computing partial derivatives of a function. In the present work, sensitivity analyses are performed to investigate the impact of different aspects in salt cavern simulations for cyclic operations. We aim to investigate the impact of model calibration, transient and reverse creep, cavern geometry, non-salt inter-layers, and mutual interaction in systems of caverns. For this purpose, an open-source three-dimensional finite element simulator has been developed to include a salt rock constitutive model that considers transient, reverse and steady-state creep. The simulator is called ``SafeInCave''.

The remainder of this manuscript is organized as follows. Section \ref{sec:mathematical_model} describes the governing equations and the constitutive model adopted to describe salt rock mechanics. The numerical formulation is presented in Section \ref{sec:numerical_formulation}, which includes description of the time integration schemes, stress linearization (consistent tangent matrix), numerical treatment of each element of the constitutive model, and the weak form for the finite element formulation. Next, Section \ref{sec:methodology} describes the different constitutive model variations adopted in our analyses, the cavern geometries, boundary conditions (pressure schedules), initial conditions, and sets of material properties used. In the results section (Section \ref{sec:results}), each analysis is appropriately described based on the information provided in Section \ref{sec:methodology}, and the obtained results are shown and discussed. Finally, the main conclusions are summarized in Section \ref{sec:conclusion}.

\section{Mathematical model}
\label{sec:mathematical_model}
\begin{figure}[!h]
	\centering
	\includegraphics[scale=0.5]{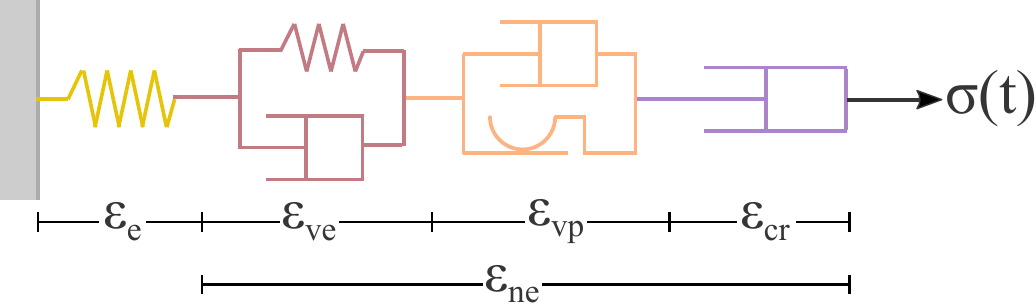}
	\caption{Elements composing the constitutive model.}
    \label{fig:fig_2}
\end{figure}

The mechanical behavior of salt caverns is described by the linear momentum equilibrium equation, which reads
\begin{equation}
    \nabla \cdot \pmb{\sigma} = \mathbf{f},
    \label{eq:mom_balance_0}
\end{equation}

\noindent where $\pmb{\sigma}$ is the stress tensor and $\mathbf{f}$ is the external force per volume vector. Constitutive laws are required to relate the stress tensor $\pmb{\sigma}$ to the total strain tensor $\pmb{\varepsilon}$. In this work, we adopt the same constitutive model as in \cite{tasinafo4782265multi}, which considers an elastic element ($\pmb{\varepsilon}_e$) for an instantaneous elastic response, a viscoelastic ($\pmb{\varepsilon}_{ve}$) element for reverse creep, a viscoplastic ($\pmb{\varepsilon}_{vp}$) element for transient creep, and a dashpot for steady-state creep due to dislocation movement. Figure \ref{fig:fig_2} illustrates the complete constitutive model, where the non-elastic strain tensor is defined as the summation of all elements except the elastic one.

\noindent In this manner, the stress tensor can be written as
\begin{equation}
    \pmb{\sigma} = \mathbb{C}_0 : \left( \pmb{\varepsilon} - \pmb{\varepsilon}_{ne} \right),
    \label{eq:stress_0}
\end{equation}

\noindent where $\mathbb{C}_0$ is the 4th-order stiffness tensor associated with the spring (elastic) element. The non-elastic strain tensor is given by the summation of all non-elastic elements, $N_{ne}$, i.e.,
\begin{equation}
    \pmb{\varepsilon}_{ne} = \sum_{i=1}^{N_{ne}} \pmb{\varepsilon}_i.
\end{equation}

It is clear that computing the total non-elastic strains $\pmb{\varepsilon}_{ne}$ for each element is required for obtaining the stress tensor in Eq. \eqref{eq:stress_0}. In the subsections below, the individual non-elastic strain rates are defined.

\subsection{Viscoelastic element}
An external stress $\pmb{\sigma}$ acting on the viscoelastic element (also called Kelvin-Voigt element) is counter-balanced by the stresses acting on its spring and dashpot, i.e.,

\begin{equation}
    \pmb{\sigma} = \underbrace{\mathbb{C}_1 : \pmb{\varepsilon}_{ve}}_{\text{spring}} + \underbrace{\eta_1 \dot{\pmb{\varepsilon}}_{ve}}_{\text{dashpot}}
    \quad \Rightarrow \quad
    \dot{\pmb{\varepsilon}}_{ve} = \frac{1}{\eta_1} \left( \pmb{\sigma} - \mathbb{C}_1 : \pmb{\varepsilon}_{ve} \right),
    \label{eq:eps_rate_ve_0}
\end{equation}

\noindent where $\mathbb{C}_1$ and $\eta_1$, respectively, denoting the 4th-order stiffness tensor and viscosity associated with the spring and dashpot of the viscoelastic element.

\subsection{Viscoplastic element}
For the viscoplastic contribution, the strain rate is computed following the non-associated formulation following Perzyina's theory, i.e.,

\begin{equation}
    \dot{\pmb{\varepsilon}}_{vp} = \mu_1 \left\langle \dfrac{ F_{vp} }{F_0} \right\rangle^{N_1} \dfrac{\partial F_{vp}}{\partial \pmb{\sigma}},
    \label{eq:eps_rate_vp}
\end{equation}

\noindent where the yield function proposed by \cite{desai1987constitutive} is employed, which reads

\begin{equation}
    F_{vp}(\pmb{\sigma}, \alpha) = J_2 - (-\alpha I_1^{n} + \gamma I_1^2) \left[ \exp{(\beta_1 I_1)} - \beta \cos(3\theta) \right]^m
    \label{eq:elem_vp_F}
\end{equation}

\noindent and the potential function is given by $Q_{vp}=F_{vp}(\pmb{\sigma}, \alpha_q)$, where $\alpha_q$ is adopted as in \cite{khaledi2016stability}. In addition, $I_1$ and $J_2$ represent the first invariant of the stress tensor and the second invariant of its deviatoric part, and $\theta$ is Lode's angle. The hardening parameter $\alpha$ obeys the following hardening rule,

\begin{equation}
    \alpha = a_1 \left[ \left( \frac{a_1}{\alpha_0} \right)^{1/\eta} + \xi \right]^{-\eta}, \quad \text{where} \quad \xi = \int_{t_0}^t \sqrt{ \dot{\pmb{\varepsilon}}_{vp} : \dot{\pmb{\varepsilon}}_{vp} } \mathrm{dt}.
    \label{eq:elem_vp_alpha}
\end{equation}

In equations \eqref{eq:eps_rate_vp}, \eqref{eq:elem_vp_F} and \eqref{eq:elem_vp_alpha} the quantities $\mu_1$, $F_0$, $N_1$, $n$, $\gamma$, $\beta_1$, $\beta$, $m$, $a_1$ and $\eta$ are all material parameters. The initial hardening parameter can be computed according to the initial stress condition $\sigma_0$ such that $F_{vp}(\pmb{\sigma}_0, \alpha_0) = 0$, i.e.,

\begin{equation}
    \alpha_0 = \gamma I_1^{2-n} - I_1^{-n} J_2 \left[ \exp\left( \beta_1 I_1 \right) - \beta \cos(3\theta) \right]^{-m}.
    \label{eq:alpha_0}
\end{equation}

\subsection{Dislocation creep element}
The dislocation creep contribution is represented by a simple power law in combination with Arrhenius law \cite{ramesh2021geomechanical}, which is written as
\begin{equation}
    \dot{\pmb{\varepsilon}}_{cr} = A \exp \left( -\frac{Q}{RT} \right) q^{n-1} \mathbf{s},
    \label{eq:eps_rate_cr}
\end{equation}

\noindent where $\mathbf{s}$ and $q$ are the deviatoric and von Mises stresses, respectively. Moreover, $A$ and $n$ are material parameters, $Q$ is the activation energy, $R$ is Boltzmann's constant and $T$ is temperature.

\section{Numerical formulation}
\label{sec:numerical_formulation}
As shown in the previous section, non-elastic strains are required to compute the stress tensor, but they also depend on the stress tensor itself and, potentially, on internal parameters. In other words, the stress tensor is a nonlinear function of total strain, which makes Eq. \eqref{eq:mom_balance_0} a nonlinear problem to be solved. In these cases, deriving a consistent tangent matrix is essential for ensuring stable solutions for implicit time integration schemes \cite{perez2001consistent}. This procedure is described below.

\subsection{Stress linearization}
Linearization of the stress field allows for solving Eq. \eqref{eq:mom_balance_0} iteratively at each time step. Therefore, at iteration $k+1$ and the current time step, one has to solve
\begin{equation}
    \nabla \cdot \pmb{\sigma}^{k+1} = \mathbf{f},
    \label{eq:mom_balance_k}
\end{equation}

\noindent where 

\begin{equation}
    \pmb{\sigma}^{k+1} = \mathbb{C}_0 : \left( \pmb{\varepsilon}^{k+1} - \pmb{\varepsilon}_{ne}^{k+1} \right).
    \label{eq:stress_notlinearized}
\end{equation}

The linearization of $\pmb{\sigma}^{k+1}$ starts by integrating Eq. \eqref{eq:stress_notlinearized} in time between $t$ and $t+\Delta t$. Employing the $\theta-$rule to integrate $\pmb{\varepsilon}^{k+1}_{i}$ leads to
\begin{equation}
    \pmb{\varepsilon}^{k+1}_{i} = \pmb{\varepsilon}^t_{i} + \phi_1 \dot{\pmb{\varepsilon}}^t_{i} + \phi_2 \dot{\pmb{\varepsilon}}^{k+1}_{i},
    \label{eq:eps_i_0}
\end{equation}

\noindent where $\phi_1 = \Delta t \theta$ and $\phi_2 = \Delta t (1 - \theta)$. In this case,  $\theta=1$, $\theta=1/2$ and $\theta=0$ provide explicit, Crank-Nicolson and fully implicit time integrations, respectively. Additionally, the superscript $t$ denotes the variables evaluated at the previous time step $t$, whereas the superscript $k+1$ indicates the current iteration of the current time step $t+\Delta t$. Naturally, superscript $k$ will indicate the previous iteration of the current time step. The superscript $t+\Delta t$ is always omitted to keep a concise notation.

In general, the non-elastic strain rate is a function of the stress tensor and an internal parameter, i.e.,

\begin{equation}
    \dot{\pmb{\varepsilon}}_{i} = \dot{\pmb{\varepsilon}}_{i}\left( \pmb{\sigma}, \alpha_i \right).
    \label{eq:strain_rate_i_0}
\end{equation}

Using Taylor series to expand Eq. \eqref{eq:strain_rate_i_0} between two consecutive iterations leads to,

\begin{equation}
    \dot{\pmb{\varepsilon}}^{k+1}_{i}
    = \dot{\pmb{\varepsilon}}^{k}_{i}
    + \frac{\partial \dot{\pmb{\varepsilon}}_{i}}{\partial \pmb{\sigma}} : \delta \pmb{\sigma}
    + \frac{\partial \dot{\pmb{\varepsilon}}_{i}}{\partial \alpha_i} \delta \alpha_i,
    \label{eq:eps_rate_i_0}
\end{equation}

\noindent where $\delta \pmb{\sigma} = \pmb{\sigma}^{k+1} - \pmb{\sigma}^k$ and $\delta \alpha_i = \alpha_i^{k+1} - \alpha_i^k$. The increment of the internal variable, $\delta \alpha_i$, can be obtained by using the evolution equation of $\alpha_i$ to define a residue function which can be generally written as

\begin{equation}
    r_i = r_i(\pmb{\sigma}, \alpha_i).
\end{equation}

\noindent Finally, using Taylor series to expand this residue function from $r_{i}^k$ to $r_{i}^{k+1}$ and applying Newton-Raphson approach yields

\begin{equation}
    \cancelto{0}{r_{i}^{k+1}} = r_{i}^k + \underbrace{\frac{\partial r_{i}^k}{\partial \alpha_{i}}}_{h_i} \delta \alpha_{i} + \frac{\partial r_{i}^k}{\partial \pmb{\sigma}} : \delta \pmb{\sigma} = 0
    \quad \Rightarrow \quad
    \delta \alpha_{i} = - \frac{1}{h_i} \left( r^k_{i} + \frac{\partial r_i^k}{\partial \pmb{\sigma}} : \delta \pmb{\sigma} \right).
    \label{eq:delta_alpha_0}
\end{equation}

By combining Equations \eqref{eq:delta_alpha_0} and \eqref{eq:eps_rate_i_0} and substituting into Eq. \eqref{eq:eps_i_0}, the strain tensor of the current iteration is expressed as

\begin{equation}
    \pmb{\varepsilon}_{i}^{k+1} = 
    \bar{\pmb{\varepsilon}}_{i}^k
    + \phi_2 \mathbb{G}_i : \delta \pmb{\sigma}
    - \phi_2 \mathbf{B}_i,
\end{equation}

\noindent where

\begin{align}
    &\bar{\pmb{\varepsilon}}_{i}^k = \pmb{\varepsilon}_{i}^t + \phi_1 \dot{\pmb{\varepsilon}}_{i}^t + \phi_2 \dot{\pmb{\varepsilon}}^{k}_{i},
    \\
    &\mathbb{G}_i = \frac{\partial \dot{\pmb{\varepsilon}}_{i}}{\partial \pmb{\sigma}} - \frac{1}{h_i} \frac{\partial \dot{\pmb{\varepsilon}}_{i}}{\partial \alpha_i} \frac{\partial r_i^k}{\partial \pmb{\sigma}}, \label{eq:G_i}
    \\
    &\mathbf{B}_i = \frac{r_i^k}{h_i} \frac{\partial \dot{\pmb{\varepsilon}}_{i}}{\partial \alpha_i}. \label{eq:B_i}
\end{align}

Finally, the non-elastic strain tensor is given by

\begin{equation}
    \pmb{\varepsilon}_{ne}^{k+1} = 
    \sum_{i=1}^{N_{ne}} \pmb{\varepsilon}_i
    =
    \bar{\pmb{\varepsilon}}_{ne}^k
    + \phi_2 \mathbb{G}_{ne} : \delta \pmb{\sigma}
    - \phi_2 \mathbf{B}_{ne},
    \label{eq:strain_ne_k_plus_1}
\end{equation}

\noindent with $\bar{\pmb{\varepsilon}}_{ne}^k = \sum_{i=1}^{N_{ne}} \bar{\pmb{\varepsilon}}_{i}^k$, $\mathbb{G}_{ne} = \sum_{i=1}^{N_{ne}} \mathbb{G}_{i}$ and $\mathbf{B}_{ne} = \sum_{i=1}^{N_{ne}} \mathbf{B}_{i}$.

From Eq. \eqref{eq:strain_ne_k_plus_1}, the linearized form of the stress tensor is expressed by

\begin{equation}
    \pmb{\sigma}^{k+1} = \mathbb{C}_T :
    \left(
        \pmb{\varepsilon}^{k+1}
        - \bar{\pmb{\varepsilon}}_{ne}^k
        + \phi_2 \mathbb{G}_{ne} : \pmb{\sigma}^k
        + \phi_2 \mathbf{B}_{ne}
    \right),
    \label{eq:stress_linearized}
\end{equation}

\noindent where $\mathbb{C}_T = \left( \mathbb{C}_0^{-1} + \phi_2 \mathbb{G}_{ne} \right)^{-1}$. Combining Equations \eqref{eq:stress_linearized} and \eqref{eq:mom_balance_k} and rearranging the terms, the linearized form of the momentum balance equation reads

\begin{equation}
    \nabla \cdot \mathbb{C}_T : \pmb{\varepsilon}^{k+1}
    =
    \mathbf{f}
    + \nabla \cdot \mathbb{C}_T : \pmb{\varepsilon}_\text{rhs}^k,
    \label{eq:mom_balance_linearized}
\end{equation}

\noindent where $\pmb{\varepsilon}_\text{rhs}^k = \bar{\pmb{\varepsilon}}_{ne}^k - \phi_2 \mathbb{G}_{ne} : \pmb{\sigma}^{k} - \phi_2 \mathbf{B}_{ne}$.

On the right-hand side of Eq. \eqref{eq:stress_linearized}, the quantities $\pmb{\varepsilon}_i$, $\dot{\pmb{\varepsilon}}_i$, $\mathbb{G}_i$ and $\mathbf{B}_i$ must be computed for all elements included in the constitutive model. The following subsections show this procedure for each element.

\subsection{Viscoelastic element}
An expression for the viscoelastic strain rate can be obtained by combining Equations \eqref{eq:eps_rate_ve_0} and \eqref{eq:eps_i_0}, with $i=ve$, and solving for $\dot{\pmb{\varepsilon}}_{ve}$. This results in the following equations

\begin{equation}
    \dot{\pmb{\varepsilon}}_{ve} = \left( \eta_1 \mathbb{I} + \phi_2 \mathbb{C}_1 \right)^{-1} : \left[ \pmb{\sigma} - \mathbb{C}_1 : \left( \pmb{\varepsilon}_{ve}^t + \phi_1 \dot{\pmb{\varepsilon}}_{ve}^t \right) \right]
    \label{eq:eps_rate_ve_1}
\end{equation}

\noindent where $\mathbb{I}$ represents the 4th-order identity tensor. The derivative of Eq. \eqref{eq:eps_rate_ve_1} with respect to $\pmb{\sigma}$ gives,

\begin{equation}
    \mathbb{G}_{ve} = \frac{\partial \dot{\pmb{\varepsilon}}_{ve}}{\partial \pmb{\sigma}} = \left( \eta_1 \mathbb{I} + \phi_2 \mathbb{C}_1 \right)^{-1}.
    \label{eq:G_ve}
\end{equation}

Furthermore, the viscoelastic model has no internal variables, implying that $\mathbf{B}_{ve} = 0$. Finally, the viscoelastic strain at iteration $k+1$ can be computed as,

\begin{equation}
    \pmb{\varepsilon}_{ve}^{k+1} = \pmb{\varepsilon}_{ve}^t + \phi_1 \dot{\pmb{\varepsilon}}_{ve}^t + \phi_2 \left( \dot{\pmb{\varepsilon}}_{ve}^k + \mathbb{G}_{ve} : \delta \pmb{\sigma} \right).
\end{equation}

\subsection{Dislocation creep element}
Equation \eqref{eq:eps_rate_cr} gives the strain rate for the dislocation creep element. The tangent matrix is given by

\begin{equation}
    \mathbb{G}_{cr} = \frac{\partial \dot{\pmb{\varepsilon}}_{cr}}{\partial \pmb{\sigma}},
\end{equation}

\noindent where the partial derivatives are computed by finite differences. Since the dislocation creep model does not depend on internal variables ($\mathbf{B}_{cr} = 0$), the creep strain at iteration $k+1$ is computed as,

\begin{equation}
    \pmb{\varepsilon}_{cr}^{k+1} = \pmb{\varepsilon}_{cr}^t + \phi_1 \dot{\pmb{\varepsilon}}_{cr}^t + \phi_2 \left( \dot{\pmb{\varepsilon}}_{cr}^k + \mathbb{G}_{cr} : \delta \pmb{\sigma} \right).
\end{equation}

\subsection{Viscoplastic element}
The viscoplastic strain rate is computed by Eq. \eqref{eq:eps_rate_vp}, where $\alpha$ is an internal variable. For this reason, a residual function is defined based on the hardening rule as

\begin{equation}
    r^k_{vp} = \alpha^k - a_1 \left[ \left( \frac{a_1}{\alpha_0} \right)^{1/\eta} + \xi^k \right]^{-\eta}, \quad \text{where} \quad \xi^k = \int_{t_0}^t \sqrt{ \dot{\pmb{\varepsilon}}_{vp}^k : \dot{\pmb{\varepsilon}}_{vp}^k } \mathrm{dt}
\end{equation}

\noindent from which $h_{vp}^k = \frac{\partial r_{vp}^k}{\partial \alpha}$, as shown in Eq. \eqref{eq:delta_alpha_0}, and Equations \eqref{eq:G_i} and \eqref{eq:B_i} are employed. All derivatives are computed by finite differences. The viscoplastic strain at iteration $k+1$ is

\begin{equation}
    \pmb{\varepsilon}_{vp}^{k+1} = \pmb{\varepsilon}_{vp}^t + \phi_1 \dot{\pmb{\varepsilon}}_{vp}^t + \phi_2 \left( \dot{\pmb{\varepsilon}}_{vp}^k + \mathbb{G}_{vp} : \delta \pmb{\sigma} \right) - \phi_2 \mathbf{B}_{vp}.
\end{equation}

\subsection{Weak formulation}
Consider a domain $\Omega$ bounded by a surface $\Gamma$ outward oriented by a unitary vector $\mathbf{n}$. Furthermore, consider a vector test function $\mathbf{v}\in \mathcal{V}$, where $\mathcal{V}$ is a test function space. The weak form of Eq. \eqref{eq:mom_balance_linearized} can then be written as

\begin{equation}
    \int_\Omega \mathbb{C}_T : \pmb{\varepsilon} \left( \mathbf{u}^{k+1} \right) : \pmb{\varepsilon} \left( \mathbf{v} \right) \text{d} \Omega
    =
    \int_\Omega \mathbf{f} \cdot \mathbf{v} \text{d} \Omega
    +
    \int_\Gamma \mathbf{t} \cdot \mathbf{v} \text{d} \Gamma
    +
    \int_\Omega \mathbb{C}_T : \pmb{\varepsilon}_\text{rhs}^k : \pmb{\varepsilon} \left( \mathbf{v} \right) \text{d} \Omega
\end{equation}

\noindent where $\mathbf{t} = \pmb{\sigma} \cdot \mathbf{n}$ and $\mathbf{u} \in \mathcal{V}$. Additionally, for any $\mathbf{w}\in\mathcal{V}$, the small strain assumption implies that

\begin{align}
    \pmb{\varepsilon}(\mathbf{w}) = \frac{1}{2} \left( \nabla \mathbf{w} + \nabla \mathbf{w}^T \right).
\end{align}

\section{Methodology}
\label{sec:methodology}
In this work, several investigations are carried out to understand the behavior of salt caverns according to different choices made during the simulation setup. The objective is to analyze the impact of considering (i) different deformation mechanisms, (ii) different calibrations of the constitutive models for the same salt rock, (iii) regular and irregular cavern shapes, (iv) the presence of a non-salt interlayer, and (v) the presence of nearby caverns (system of caverns). In this section, we describe the constitutive models, material parameters, geometries, boundary/initial conditions, and laboratory experiments adopted in this work. The descriptions of each test case are left for the results section (Section \ref{sec:results}).

\subsection{Constitutive models}
\label{subsec:constitutive_models}
\begin{figure}[!h]
	\centering
	\includegraphics[scale=0.25]{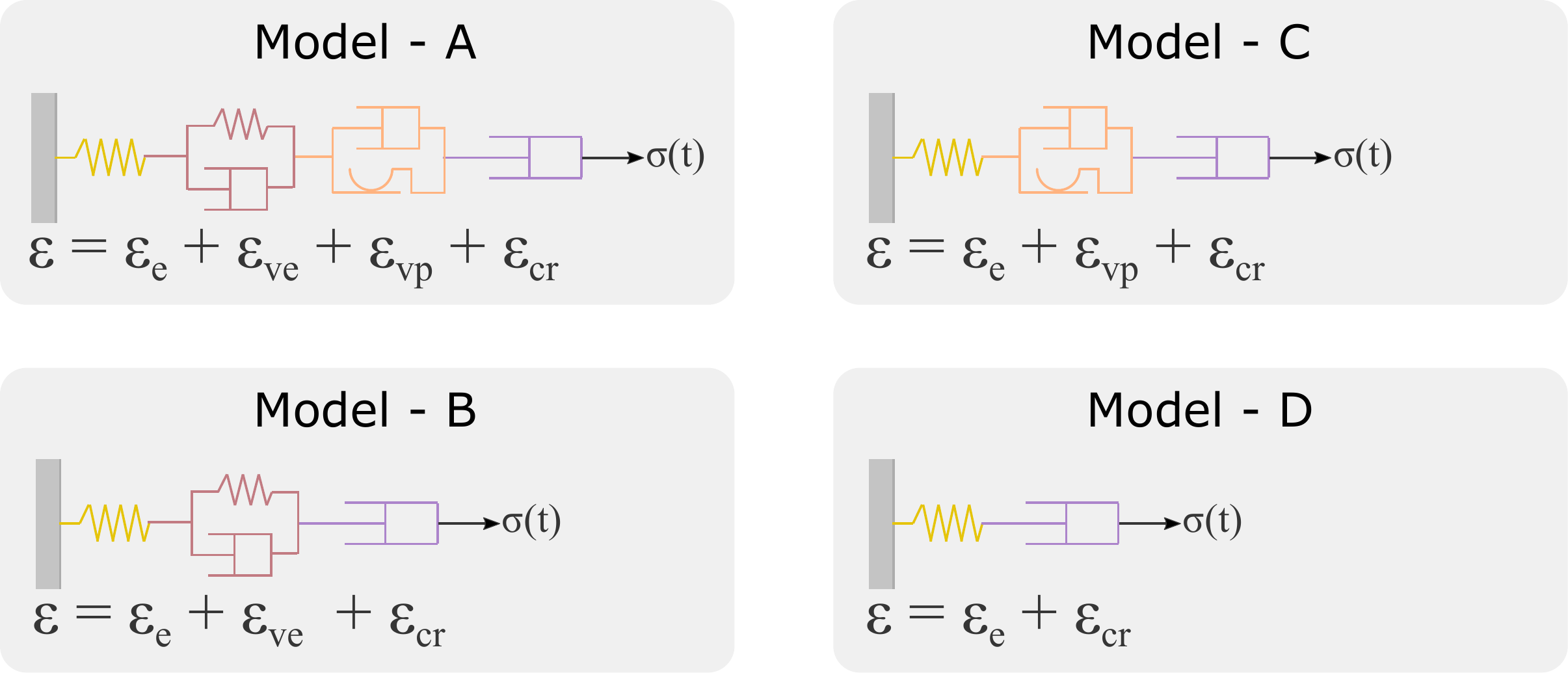}
	\caption{Constitutive models A (elastic, viscoelastic, viscoplastic and dislocation creep), B (elastic, viscoelastic and dislocation creep), C (elastic, viscoplastic and dislocation creep), and D (elastic and dislocation creep).}
    \label{fig:fig_3}
\end{figure}

The constitutive models considered in this work are depicted in Figure \ref{fig:fig_3}. Model-A includes all elements discussed in Sections \ref{sec:mathematical_model} and \ref{sec:numerical_formulation}. The viscoplastic and viscoelastic elements are removed from models B and C, respectively. Finally, Model-D only considers elastic and dislocation creep. The comparison of these constitutive models is used to assess the importance of viscoelasticity and viscoplasticity on the mechanical behavior of salt rocks.

\subsection{Cavern geometries}
\begin{figure}[!h]
	\centering
	\includegraphics[scale=0.25]{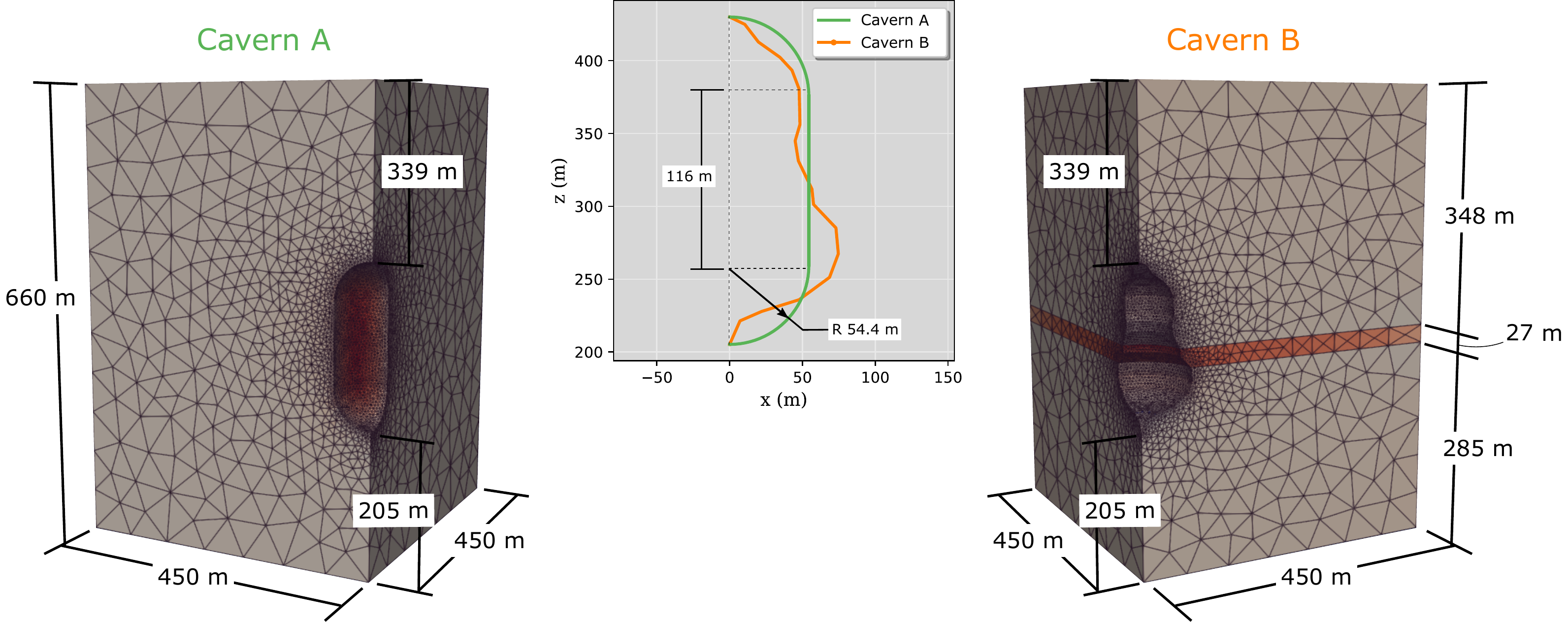}
	\caption{Single cavern geometries with different cavern shapes (A and B).}
    \label{fig:fig_4}
\end{figure}

Irregular cavern geometries create regions of stress concentrations that can induce more creep and increase cavern closure. To measure this effect, we consider two different cavern shapes, as shown in Fig. \ref{fig:fig_4}. Cavern A has a hypothetical regular shape with smooth curvatures and no sharp angles. On the other hand, cavern B is meant to reproduce a more realistic scenario, where a complex geometry wider at the bottom and narrower towards the top is adopted. For the sake of comparison, both caverns have the same volume and height. 

The impact of a non-salt layer intercepting the cavern is investigated using cavern B. It can be noticed in Fig. \ref{fig:fig_4}, that cavern B has a region that can represent a non-salt layer or salt, depending on the material parameters adopted.

\subsection{Boundary conditions}
\begin{figure}[!h]
	\centering
	\includegraphics[scale=0.3]{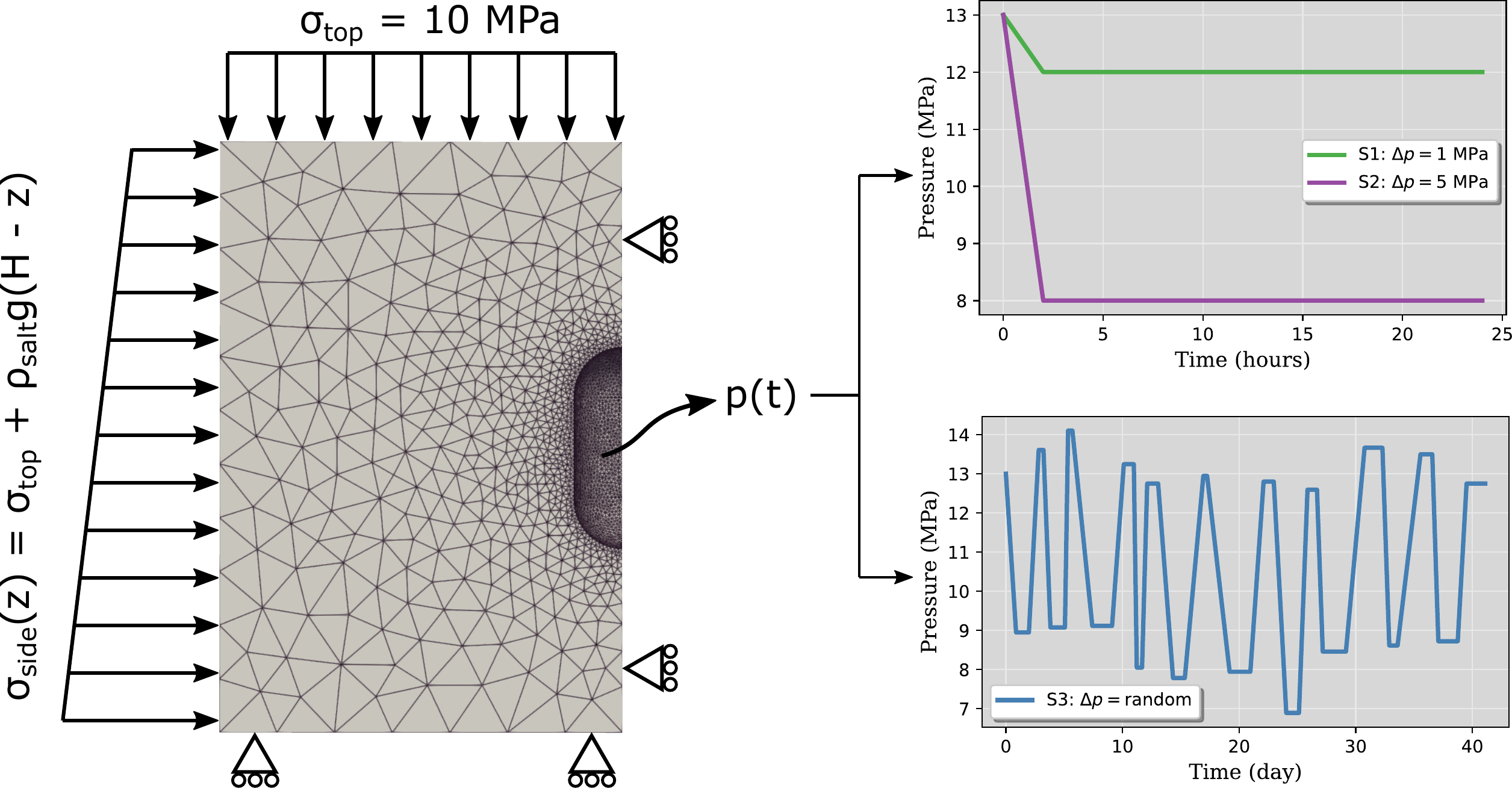}
	\caption{Boundary conditions and pressure schedules (i.e., operational fluctuating pressure values) for regular depletion (top-right) and random cyclic (bottom-right) scenarios.}
    \label{fig:fig_5}
\end{figure}

The boundary conditions applied to the geometries are depicted in Fig. \ref{fig:fig_5}. The overburden is included by imposing a load of 10 MPa on the top boundary. The two vertical plane boundaries not in contact with the cavern are subjected to a side burden following the lithostatic pressure with $\rho_\text{salt} = 2000$ kg$/$m$^3$. The other two vertical plane boundaries are prevented from normal displacement. The gas pressure is imposed as a uniform pressure distribution on the cavern walls, and the cavern pressure schedules are shown in Fig. \ref{fig:fig_5}. Pressure schedules S1 and S2 start from 13 MPa followed by 2 hours of production, during which the cavern pressure drops by 1 MPa and 5 MPa, respectively. Compared to S2, the pressure schedule S1 is expected to cause less stresses on the cavern walls. Both schedules S1 and S2 have a short duration of 24 hours. On the other hand, the pressure schedule S3, shown below, represents a more realistic case in which fast and irregular cyclic operation is imposed for approximately 40 days.

\begin{figure}[!h]
	\centering
	\includegraphics[scale=0.15]{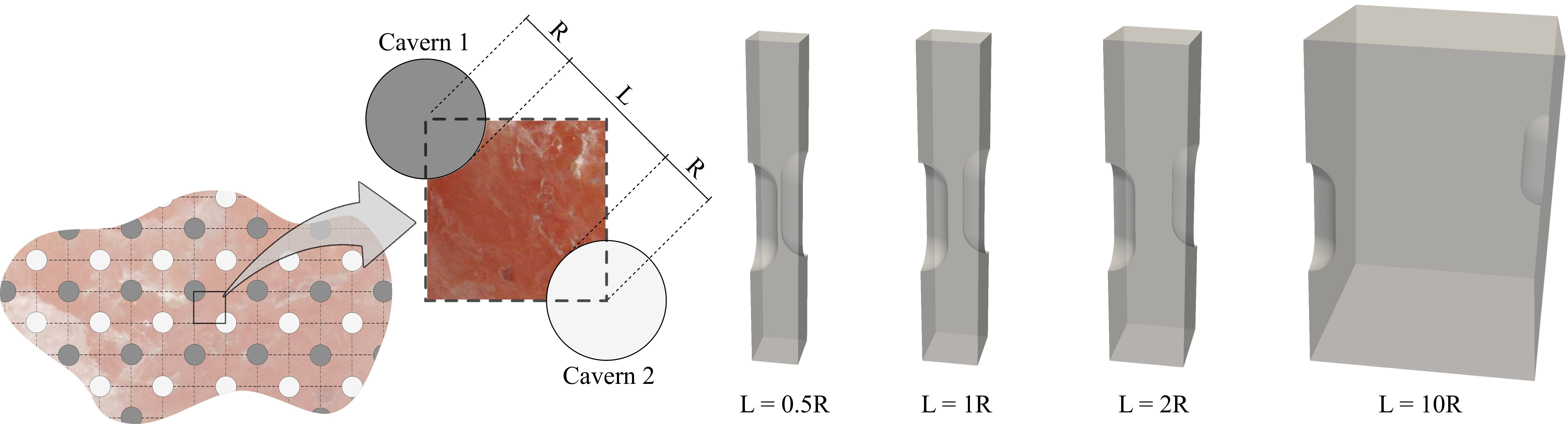}
	\caption{System of caverns configuration and geometries employed.}
    \label{fig:fig_6}
\end{figure}

\begin{figure}[!b]
	\centering
	\includegraphics[scale=0.45]{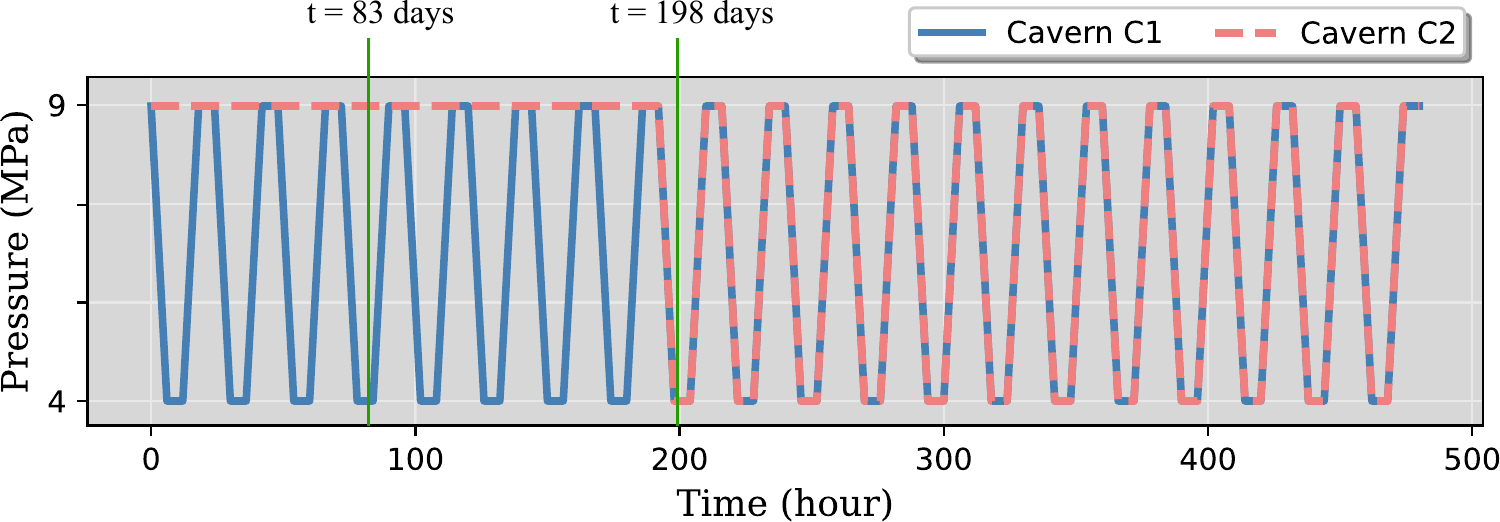}
	\caption{Pressure schedule S4 imposed to caverns 1 and 2, shown in Fig. \ref{fig:fig_6}.}
    \label{fig:fig_7}
\end{figure}

In addition to single cavern simulations, we also investigate the mechanical behavior of systems of caverns. In this work, we consider the caverns to be equally spaced from each other and placed in alternate positions as shown in Fig. \ref{fig:fig_6}. The symmetries associated to this configuration allows for the simulation of only a small portion of the system, in a similar manner as the well-known five spot problem for reservoir simulations. All vertical planar boundaries are prevented from normal displacement. Additionally, the chosen distances between the caverns (pillar width) are $L=0.5R$, $L=1R$, $L=2R$ and $L=10R$, as shown in Fig. \ref{fig:fig_6}. The pressure schedules imposed to Caverns 1 and 2 are illustrated in Fig. \ref{fig:fig_7}. As shown in this figure, $R$ is the characteristic radius of the caverns. 

\subsection{Initial condition}
The initial condition is computed by solving an equilibrium problem before the pressure schedule is applied. In the equilibrium stage, a constant pressure of $p(t=0) = 13$ MPa (for all S1, S2 and S3 scenarios) with the boundary conditions as illustrated in Fig. \ref{fig:fig_5} until the steady-state (equilibrium) condition is reached. In addition, only the elastic and viscoelastic elements are considered for the equilibrium condition. By the end of this stage, the equilibrium viscoelastic strain and strain rate are assigned to the initial conditions of the operation stage, where the pressure schedule is imposed on the cavern. The initial strain and strain rates are considered zero for the inelastic elements.

After the equilibrium state is obtained, the resulting stress condition on every element is considered to be located at the onset of viscoplasticity, i.e., on the yield surface. Therefore, any additional load applied to the cavern will cause viscoplastic deformation. This is ensured by manipulating Eq. \eqref{eq:elem_vp_F} with $F_{vp}(\pmb{\sigma}, \alpha_0) = 0$ and solving for the initial hardening parameter, i.e.,

\begin{equation}
    \alpha_0 = \gamma I_1^{2-n} - \frac{J_2}{I_1^n} \left( \exp{\left( \beta_1 I_1 \right)} + \beta \cos{(3\theta)} \right).
\end{equation}

\subsection{Factor of Safety}
\label{subsec:fos}
The Factor of Safety (FOS) is a practical quantity often used to assess the likelihood of tertiary creep to occur \cite{khaledi2016stability}. It is defined as

\begin{equation}
    \text{FOS} = \frac{\sqrt{F_\text{dil}(\pmb{\sigma})}}{\sqrt{J_2}},
\end{equation}

\noindent where $F_\text{dil}(\pmb{\sigma})$ describes the compressibility/dilatancy boundary according to Desai's \cite{desai1987constitutive} model, and it is given by

\begin{equation}
    F_\text{dil}(\pmb{\sigma}) = \left( 1 - \frac{2}{n} \right) \gamma I_1^2 \left[ \exp \left( \beta_1 I_1 \right) - \beta \cos \left( 3 \theta \right) \right].
\end{equation}

\noindent Tertiary creep is expected to occur when FOS$\leq 1.0$.

\subsection{Laboratory experiment}
\label{subsec:lab_exp}
\begin{figure}[!h]
	\centering
	\includegraphics[scale=0.6]{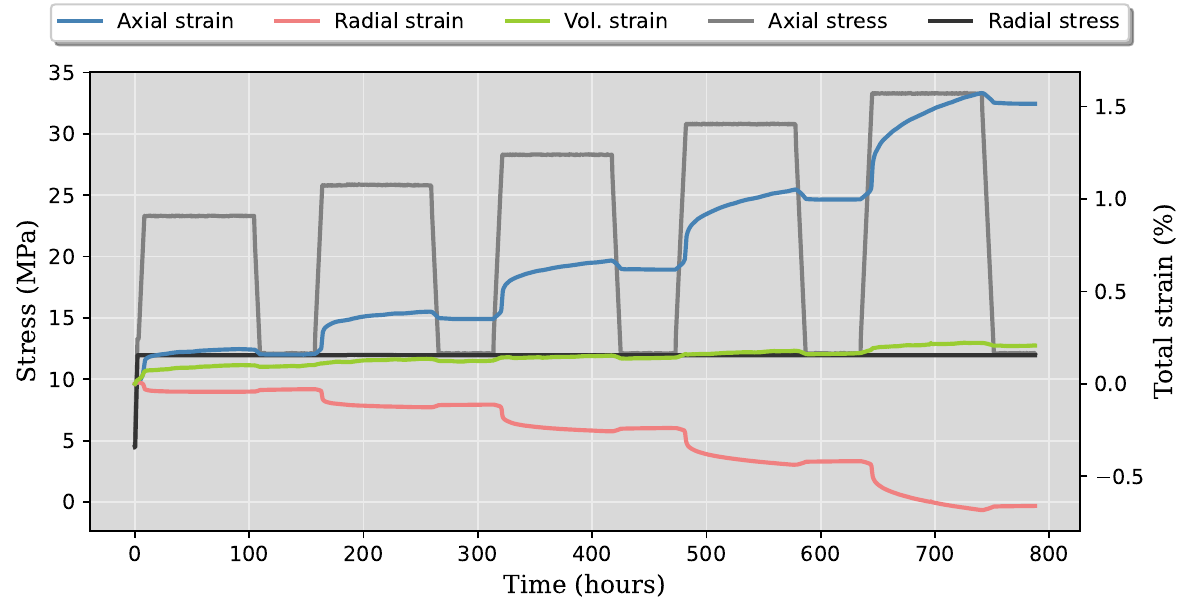}
	\caption{Experimental results obtained from a cyclic loading triaxial test. Extracted from \cite{tasinafo4782265multi}.}
    \label{fig:fig_8}
\end{figure}

The laboratory experiment used in this paper consists of a triaxial test on a salt sample, where the axial stress is applied cyclically and the confining pressure (radial stress) is kept at a constant level. The stress conditions and the axial and radial strain measured during the experiment are shown in Fig. \ref{fig:fig_8}. The details of the experimental setup can be found in \cite{tasinafo4782265multi}.

\subsection{Material parameters}
Table \ref{tab:mat_params} shows four material parameter sets for the constitutive model: Salt-A, Salt-B, Anhydrite and Mudstone. The Salt-A and Salt-B are possible parameter sets that provide reasonably good fitting against the laboratory experiment described in subsection \ref{subsec:lab_exp}. The anhydrite and mudstone sets represent the interlayer in Cavern B (see Fig. \ref{fig:fig_4}). This layer is assumed to be purely elastic, with no time-dependent behavior. The elastic properties of anhydrite and mudstone are taken from \cite{hangx2010mechanical} and \cite{zhao2023creep}, respectively.

\begin{table}[!h]
\centering
\caption{Sets of material properties for halite (Salt-A, Salt-B), anhydrite and mudstone.}
\begin{tabular}{cclcccc}
Element                            & Property        & Units                     & Salt-A   & Salt-B   & Anhydrite & Mudstone \\[0.5em] \hline \\[-0.6em]
\multirow{2}{*}{Elastic}           & $E_0$           & GPa                       & 79       & 118      & 61.5      & 19.33     \\
                                   & $\nu_0$         & -                         & 0.32     & 0.32     & 0.32      & 0.223      \\[0.7em]
\multirow{3}{*}{Viscoelastic}      & $E_1$           & GPa                       & 45       & 42       & -         & -         \\
                                   & $\nu_1$         & -                         & 0.32     & 0.32     & -         & -         \\
                                   & $\eta_1$        & Pa.s                      & 3.7e14   & 2.5e14   & -         & -         \\[0.7em]
\multirow{11}{*}{Viscoplastic}     & $\mu_1$         & s$^{-1}$                  & 1e-12    & 6.65e-13 & -         & -         \\
                                   & $N_1$           &  -                        & 3.053    & 3.053    & -         & -         \\
                                   & $a_1$           & MPa$^{2-n}$               & 1.3e-05  & 2e-05    & -         & -         \\
                                   & $\eta$          &  -                        & 0.827    & 0.8      & -         & -         \\
                                   & $\beta_1$       & MPa$^{-1}$                & 0.004459 & 0.001011 & -         & -         \\
                                   & $\beta$         &   -                       & 0.995    & 0.995    & -         & -         \\
                                   & $m$             &  -                        & -0.5     & -0.5     & -         & -         \\
                                   & $n_1$           &  -                        & 3.0      & 3.0      & -         & -         \\
                                   & $\gamma$        &   -                       & 0.088012 & 0.088012 & -         & -         \\
                                   & $k$             &  -                        & 0.268738 & 0.33     & -         & -         \\
                                   & $\sigma_t$      & MPa                       & 5.4      & 5.4      & -         & -         \\[0.7em]
\multirow{5}{*}{Dislocation creep} & $A$             & Pa$^{-n}$s$^{-1}$         & 5.9e-29  & 1.1e-21  & -         & -         \\
                                   & $n$             & -                         & 4.0      & 3.0      & -         & -         \\
                                   & $T$             & K                         & 298      & 298      & -         & -         \\
                                   & $Q$             & J/mol                     & 51600    & 51600    & -         & -         \\
                                   & $R$             & J.K$^{-1}$.mol$^{-1}$     & 8.32     & 8.32     & -       & -      
\end{tabular}
\label{tab:mat_params}
\end{table}

\subsection{Code implementation}
The three-dimensional salt cavern simulator is developed using Python language. The weak formulation presented in Section \ref{sec:numerical_formulation} is solved by finite elements using Dolfin \cite{logg2010dolfin}, from FEniCS project version 2019.1 \cite{alnaes2015fenics,logg2012automated}. The geometries and meshes are created in Gmsh \cite{geuzaine2009gmsh} version 4.10.1 and subsequently converted to .xml format using dolfin-convert command. Regarding time integration, the Crank-Nicolson ($\theta=0.5$) is employed most of the time, except when the time step sizes are required to be too small, in which case the explicit formulation ($\theta=1.0$) is adopted. Finally, the resulting linear systems are solved using PETSc \cite{dalcin2011parallel} implementation of the conjugate gradient (CG) method and the successive over-relaxation (SOR) as a preconditioner. The simulator is fully open-source, and it is available at our \href{https://gitlab.tudelft.nl/ADMIRE_Public/safeincave }{Gitlab repository}.

\section{Simulation results}
\label{sec:results}
In this section, we study the sensitivity of salt cavern deformation and the state of stress to different variables, as discussed in Section \ref{sec:methodology}.

\subsection{Influence of model calibration}
\label{subsec:model_cal_pred}

\begin{figure}[!h]
	\centering
	\includegraphics[scale=0.6]{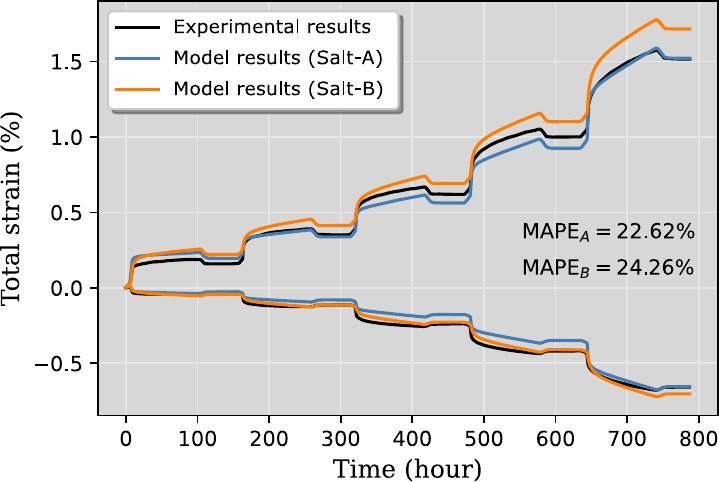}
	\caption{Model calibrations against experiments A and B. Quantities MAPE$_A$ and MAPE$_B$ represent the mean absolute percentage error for parameter sets Salt-A and Salt-B, respectively.}
    \label{fig:fig_9}
\end{figure}

In this subsection, we investigate the impact of model calibration on describing laboratory experiments and salt cavern simulations. As discussed in Section \ref{sec:introduction}, the complex mechanical behavior of salt rocks often leads to constitutive models that depend on many parameters, which renders the calibration challenging. For instance, the creep element requires that salt samples reach steady-state creep to properly define the power law coefficients. However, it is not possible to ensure the experiment shown in Fig. \ref{fig:fig_8} actually reaches this condition. Additionally, the viscoplastic element requires at least six experiments specifically designed to determine its material parameters. The elastic and viscoelastic properties are also challenging to be unambiguously defined. Consequently, there might be many combinations of material parameters that can fit a single laboratory experiment \cite{tasinafo4782265multi}. The set of parameters Salt-A and Salt-B, shown in Table \ref{tab:mat_params}, were chosen to expose this particular issue. As shown in Fig. \ref{fig:fig_9}, both parameter sets provide reasonably good description of the experimental results, with a mean absolute percentage error (MAPE) of 22.62\% and 24.26\% for Salt-A and Salt-B, respectively.

\begin{figure}[!b]
	\centering
	\includegraphics[scale=0.6]{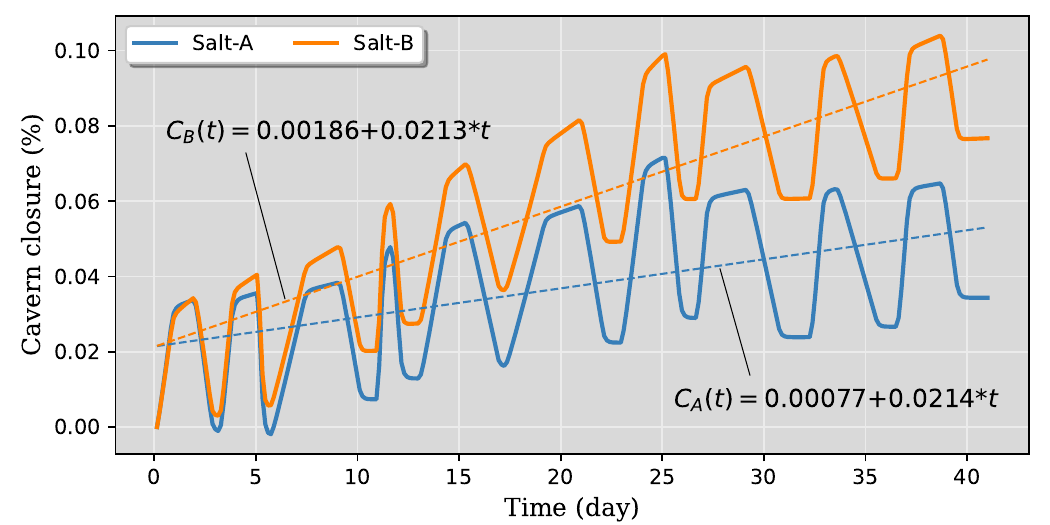}
	\caption{Cavern volume losses obtained with Salt-A and Salt-B parameter sets.}
    \label{fig:fig_10}
\end{figure}

The fact that parameters Salt-A and Salt-B provide similar results in laboratory experiments could suggest that running a salt cavern simulation with either of them would also produce similar results. To investigate this hypothesis, salt cavern simulations are performed considering both material parameter set: Salt-A and Salt-B. In this case, the Cavern A of Fig. \ref{fig:fig_4} is employed and the pressure schedule S3 (Fig. \ref{fig:fig_5}) is imposed on the cavern walls. The volume loss of the cavern is monitored over time and the results are presented in Fig. \ref{fig:fig_10}. After 40 days of operation, the difference between the two simulations is substantial. Two linear equations, $C_A(t)$ and $C_B(t)$, are fitted against both solutions using least-squares. Using these equations to predict the cavern closure after 100 years of operation provide around 28\% and 68\% for Salt-A and Salt-B, respectively. This is a significant difference of results and it emphasizes the importance of model calibration for salt cavern simulations.

\subsection{Importance of deformation mechanisms}
As discussed before, the constitutive model adopted in this work includes different deformation mechanisms to capture the mechanical behavior of salt rocks. The elastic component is intended to capture the instantaneous elastic material response. Reverse creep, on the other hand, is often described as a time-dependent deformation response that follows after an unloading phase. In the stress-strain graph, this phenomenon manifests a hysteretic curve during unloading and reloading paths. It is well-known that viscoelastic materials present this behavior, which is why a Kelvin-Voigt element is included in the constitutive model. Transient creep is intended to be captured by the viscoplastic model, as presented in Section \ref{sec:mathematical_model}. Finally, dislocation creep is described by a well-established power law function. For hydrogen operations, the loading conditions are expected to constantly change during the cavern's life cycle. Therefore, transient and reverse creep might always be present, and they should be appropriately considered. The constitutive models presented in subsection \ref{subsec:constitutive_models} are employed to investigate the importance of each deformation mechanism in laboratory experiments and salt cavern simulations.

\begin{figure}[!h]
	\centering
	\includegraphics[scale=0.6]{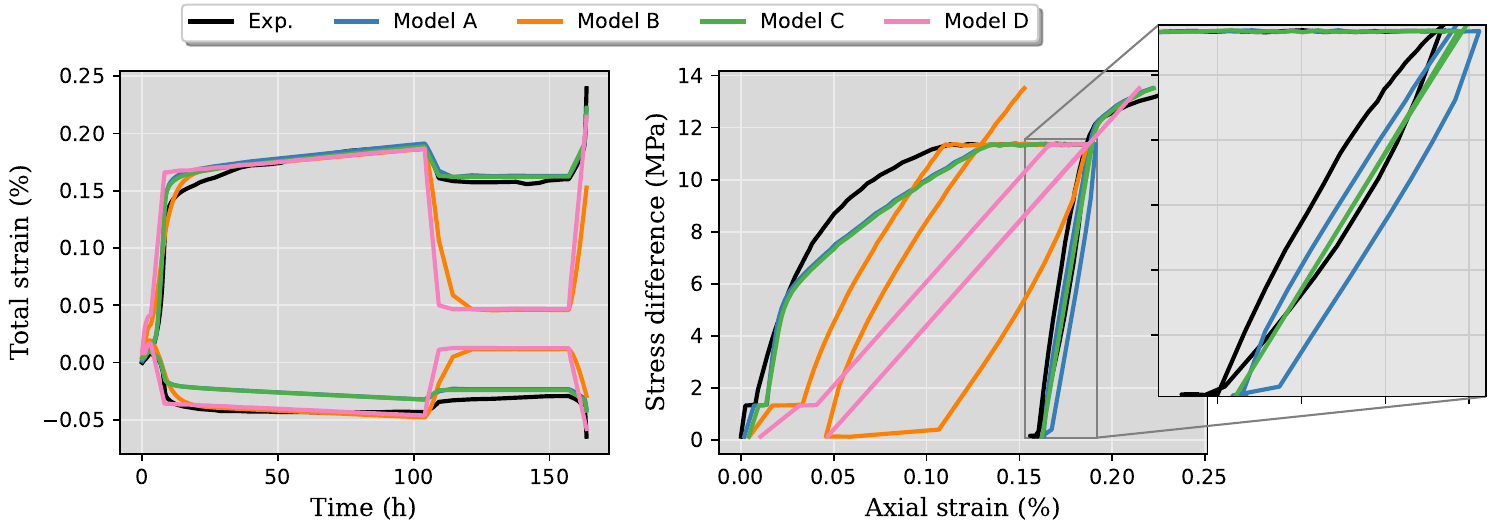}
	\caption{Results obtained with different constitutive models to describe experiment A. The models A, B, C, and D are illustrated in Fig. \ref{fig:fig_3}.}
    \label{fig:fig_11}
\end{figure}

Figure \ref{fig:fig_11} shows the first 175 hours of experiment A (see subsection \ref{subsec:lab_exp}) and the fitting results obtained with models A, B, C and D (see subsection \ref{subsec:constitutive_models}). It can be noticed that models B and D provide relatively good results during the first loading step (i.e. from 0 to 100 hours), but most of the deformation is recovered during the unloading step, which does not agree with the experimental results. The right graph of Fig. \ref{fig:fig_11} shows that the stress-strain curves obtained with these two models are completely different from the experimental data. Conversely, the inclusion of the viscoplastic element allows for models A and C to correctly describe both loading and unloading stages. However, when the stress-strain curve is considered, model C is not able to capture the reverse creep characterized by the hysteretic path observed during unloading/loading steps.

\begin{figure}[!b]
	\centering
	\includegraphics[scale=0.6]{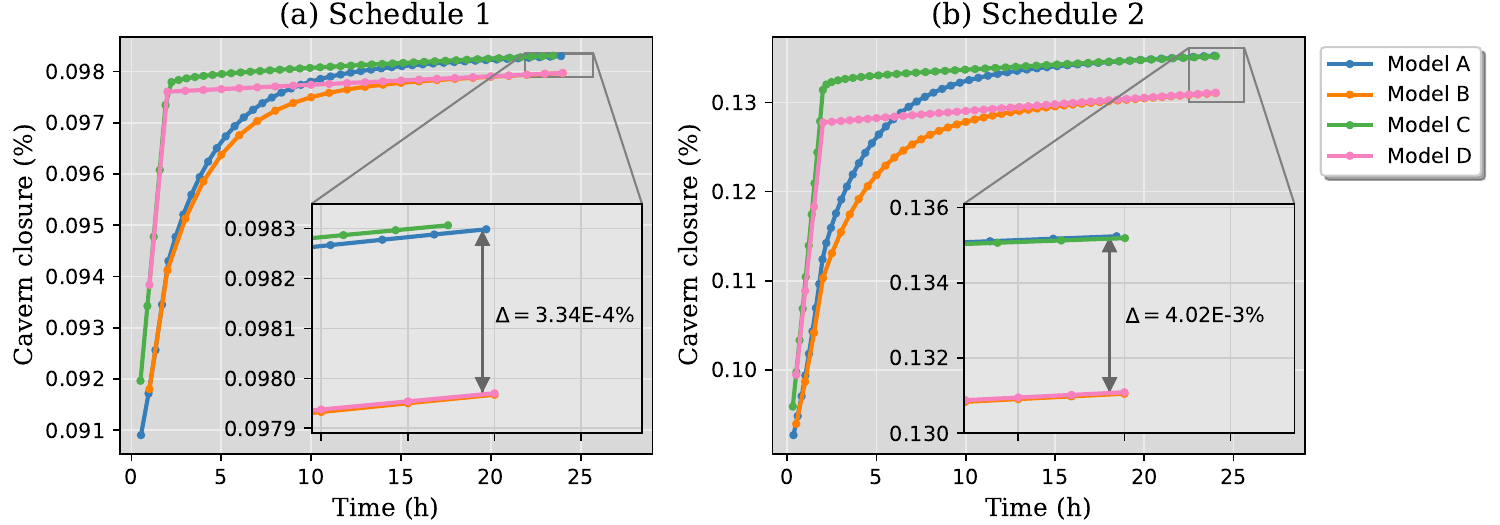}
	\caption{Cavern closure obtained with different constitutive models considering pressure schedules S1 (a) and S2 (b). The models A, B, C, and D are illustrated in Fig. \ref{fig:fig_3}.}
    \label{fig:fig_12}
\end{figure}

The same models presented in subsection \ref{subsec:constitutive_models} are now employed to solve the salt cavern problem. For this test case, the pressure schedules S1 and S2 (see Fig. \ref{fig:fig_5}) are imposed on the walls of cavern A in Fig. \ref{fig:fig_4}. The results presented in Fig. \ref{fig:fig_12}-a show a small difference between models A and B, suggesting that viscoplasticity does not make much difference. However, a significant difference between models A and C (and also between B and D) is observed, which is due to the presence or not of the viscoelastic element. For the pressure schedule S2, shown in Fig. \ref{fig:fig_12}-b, the viscoplastic contribution during gas production is increased by one order of magnitude, which can be verified by the difference between models A and B and between models C and D. The same observations regarding the viscoelastic contribution for pressure schedule S1 also hold for S2. It should be stressed, however, that both schedules S1 and S2 are inducing viscoplasticity because gas pressure is being depleted below the historical minimum, which is 13 MPa in this case. When the gas pressure increases again, additional viscoplasticity will occur only in the next production period if gas pressure drops below the minimum historical pressures (i.e. 12 MPa for S1 and 8 MPa for S2).

\begin{figure}[!h]
	\centering
	\includegraphics[scale=0.6]{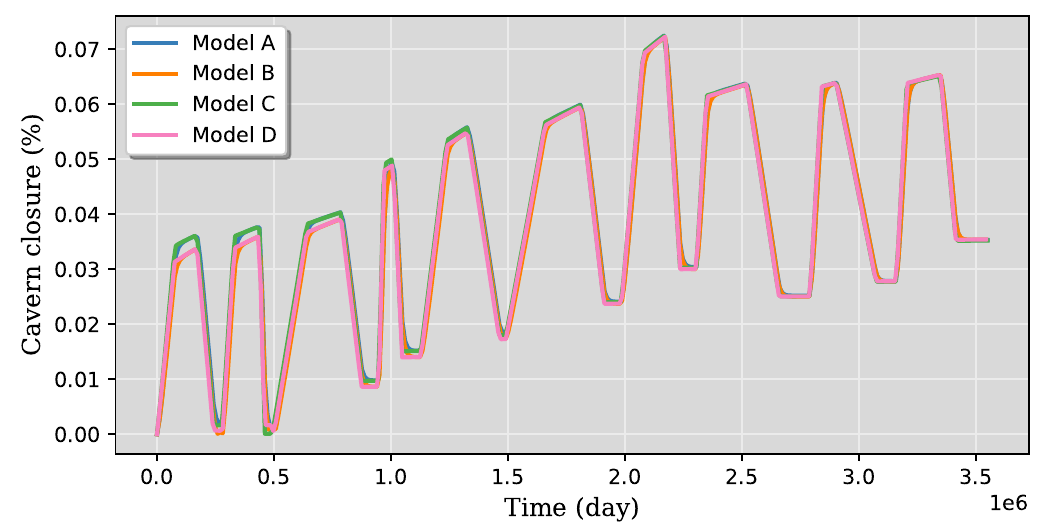}
	\caption{Cavern closure obtained with different constitutive models considering pressure schedule S3.}
    \label{fig:fig_13}
\end{figure}

The results presented in Fig. \ref{fig:fig_12} suggest that the difference between models A and B (or C and D) is expected to grow with time only if the minimum historical pressure is exceeded in the future, which would induce additional viscoplastic deformation. This situation is reproduced in pressure schedule S3, where the minimum gas pressure is often exceeded during the operation (see Fig. \ref{fig:fig_5}). To investigate this hypothesis, the pressure schedule S3 is imposed on Cavern A (Fig. \ref{fig:fig_4}) and the results are shown in Fig. \ref{fig:fig_13} for models A, B, C and D. As observed before in Fig. \ref{fig:fig_12}, model A provides similar results as model C (the same holds for models B and D), which is directly linked with the presence of the viscoplastic element. During the first three to five cycles, the two models that include the viscoplastic element (models A and C) predict higher cavern closure than models B and D, which neglect viscoplasticity. Interestingly, this difference tends to vanish as the operation progresses and all models predict virtually the same cavern closures.

\begin{figure}[!b]
	\centering
	\includegraphics[scale=0.53]{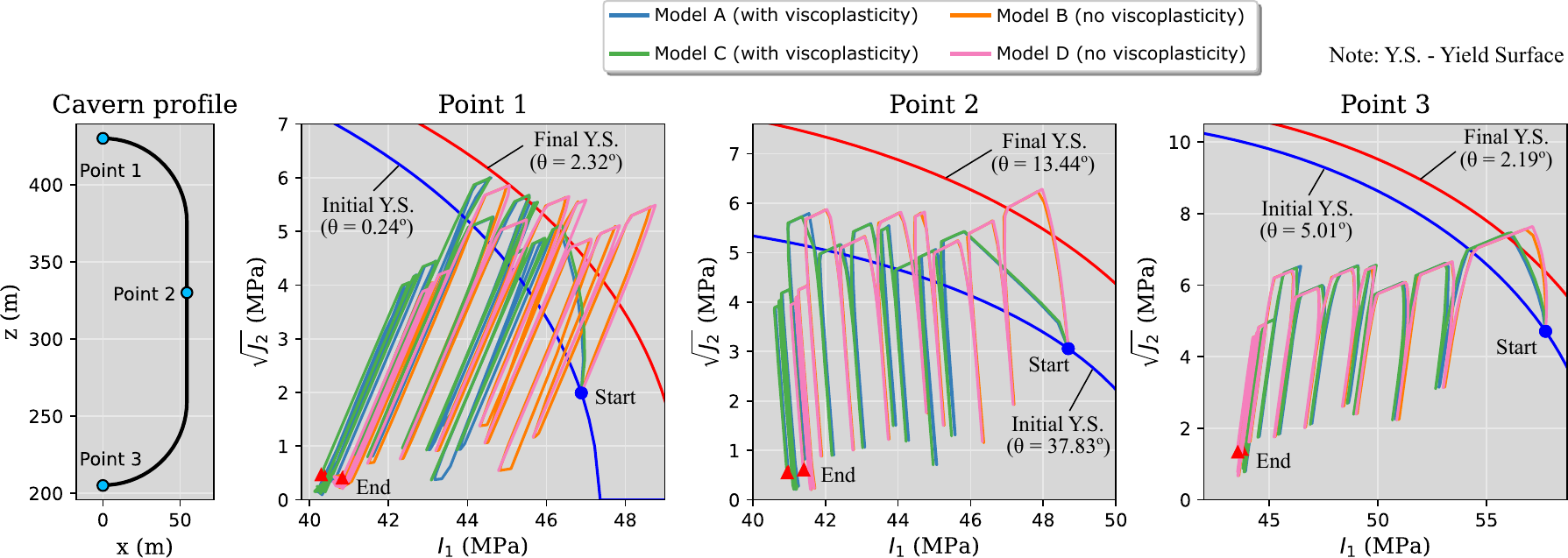}
	\caption{Stress paths at different positions of the cavern wall obtained with models A, B, C and D. The figures show the start (blue circles) and end (red triangles) points for all stress paths. The initial and final yield surfaces are also shown (note the different Lode's angle, $\theta$, for each case).}
    \label{fig:fig_14}
\end{figure}

Even though the results of Fig. \ref{fig:fig_13} suggest that the choice of the model does not significantly affect the cavern deformations, it is important to investigate the stresses around the cavern, as it can generate tertiary creep or even short-term failure. For this purpose, Fig. \ref{fig:fig_14} shows the stress paths at the top (point 1), middle (point 2), and bottom (point 3) of the cavern wall. 
Regardless of the model, all points start from the same stress state but finish at slightly different positions.
It is shown that the presence of the Kelvin-Voigt element (viscoelasticity) does not produce a significant difference, since similar results are obtained with models A-C and B-D. On the other hand, the presence of a viscoplastic element does change the stress paths, particularly during the first pressure drop inside the cavern. Figure \ref{fig:fig_14} also shows the initial and final yield surfaces for all three points. It can be verified that the three initial yield surfaces are crossed as soon as the cavern pressure starts to decrease, thus inducing viscoplastic deformation. However, the stress paths tend to move to the left (towards lower mean stresses), so additional viscoplasticity does not necessarily occur, even though the minimum gas pressure is exceeded a few times during the operation. This shows the complexity of the stress behavior around the cavern walls. Finally, it should be noted that the comparison between the initial and final position of yield surfaces must be done keeping in mind the value of the Lode's angle, $\theta$ (see Eq. \ref{eq:elem_vp_F}). This is the reason, for example, that the final yield surface of point 2 seems to be disconnected from the stress path\footnote{It can be seen that the stress paths of points 1 and 3 are touching their corresponding final yield surfaces, but this is not the case for point 2.}. 

\subsection{Impact of cavern shape}
\begin{figure}[!h]
	\centering
	\includegraphics[scale=0.14]{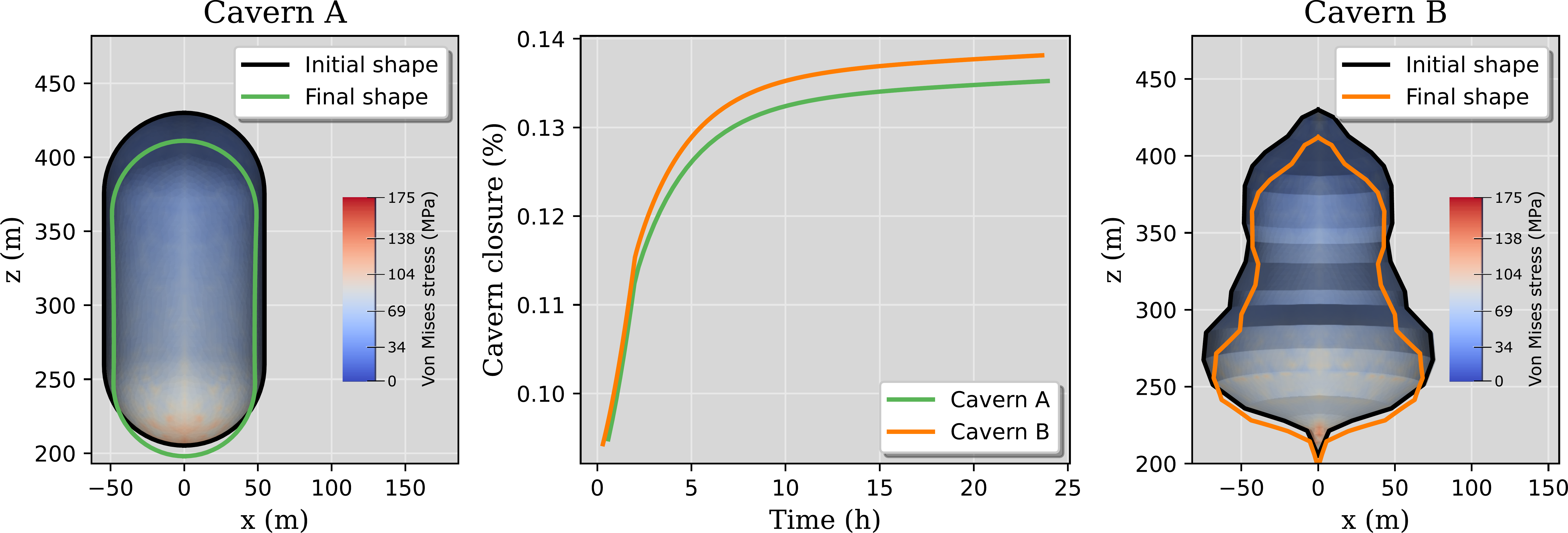}
	\caption{Initial and final shape of caverns A and B on the left and right, respectively (displacements are amplified 200 times for visualization). In the middle, cavern closure over time for both caverns.}
    \label{fig:fig_15}
\end{figure}

During cavern construction, the leaching process of the salt rock results in caverns with irregular and complex shapes. The presence of sharp angles on the cavern walls can create regions of stress concentration which induce higher creep rates and, potentially, tertiary creep. To study this case, we first compare the behavior of Cavern A (regular shape) and Cavern B (irregular shape) under the pressure schedule S2 (see Fig. \ref{fig:fig_5}). The results are presented in Fig. \ref{fig:fig_15}, which shows the initial and final shape of both caverns and the cavern closure over time. The left and right graphs also show the von Mises stresses on the cavern walls by the end of the simulation. Although it is not possible to visually identify regions of stress concentrations in Cavern B, the graph in the middle shows a higher volume loss for this cavern when compared to Cavern A. Interestingly, the difference between the two curves is observed to increase right after the production phase, that is, after the first 2 hours. This is attributed to slightly higher stresses that increase strain rates.

\begin{figure}[!t]
	\centering
	\includegraphics[scale=0.6]{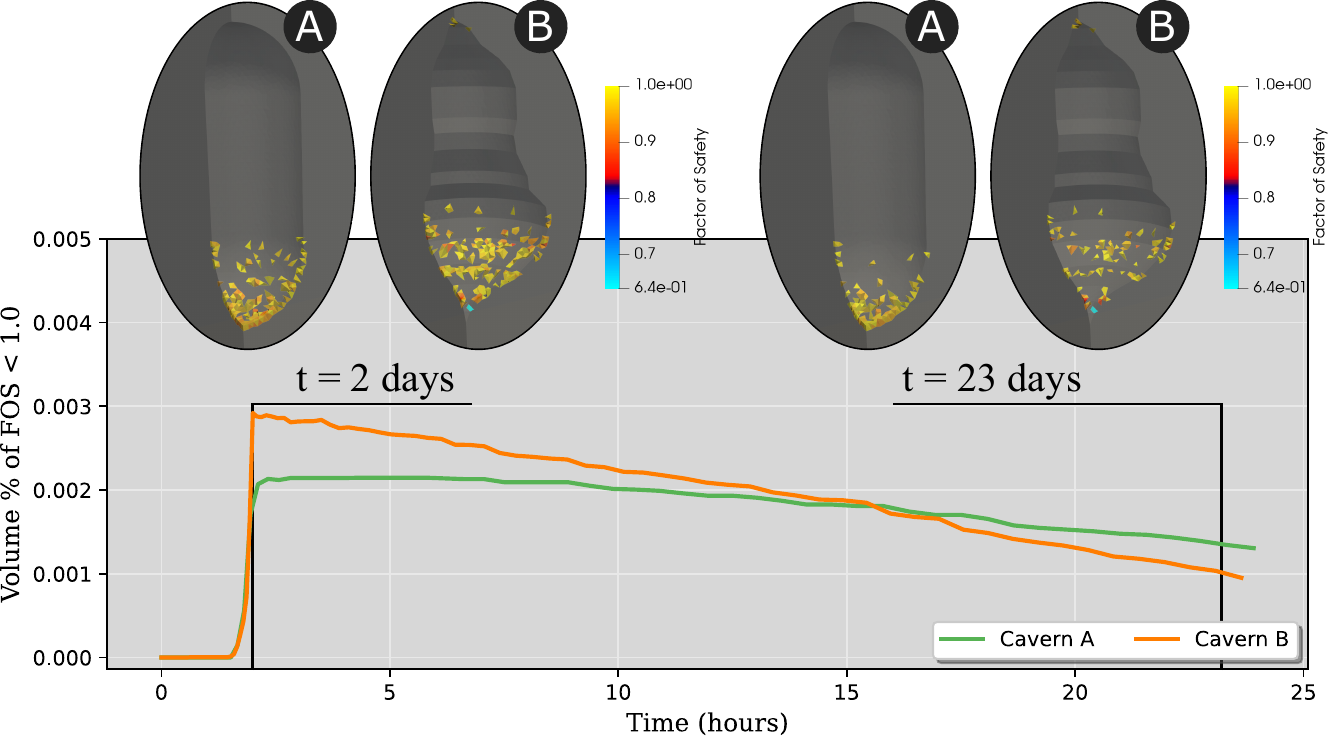}
	\caption{Percentage of salt rock volume undergoing tertiary creep (FOS$\leq1.0$).}
    \label{fig:fig_16}
\end{figure}

The possibility of tertiary creep is investigated by analyzing the factor of safety (FOS), as defined in subsection \ref{subsec:fos}. The factor of safety is computed by post-processing the stress field at each element of the grid. In the sequence, we sum up the volumes of all the elements where FOS$\leq1.0$ and divide by the total volume of the entire geometry. This gives the volumetric percentage of elements undergoing tertiary creep, and the results are presented in Fig. \ref{fig:fig_16}. By the end of the production period, the volume of salt undergoing tertiary creep (FOS$\leq1.0$) is bigger for Cavern B than for Cavern A, which is a consequence of the higher stress concentration regions. Moreover, in both cases, the volume of elements with FOS$\leq1.0$ decreases over time due to stress relaxation, but the rate of decay is higher for Cavern B. This seems to suggest that irregular cavern shapes initially induce higher stresses, thus causing more cavern closure and tertiary creep, but it also favors stress redistribution. Figure \ref{fig:fig_16} also shows the elements with FOS$\leq1.0$ for both caverns at two different times. Most of these elements are located at the bottom of the caverns, but a few elements also appear at the top of Cavern B. 

\begin{figure}[!b]
	\centering
	\includegraphics[scale=0.65]{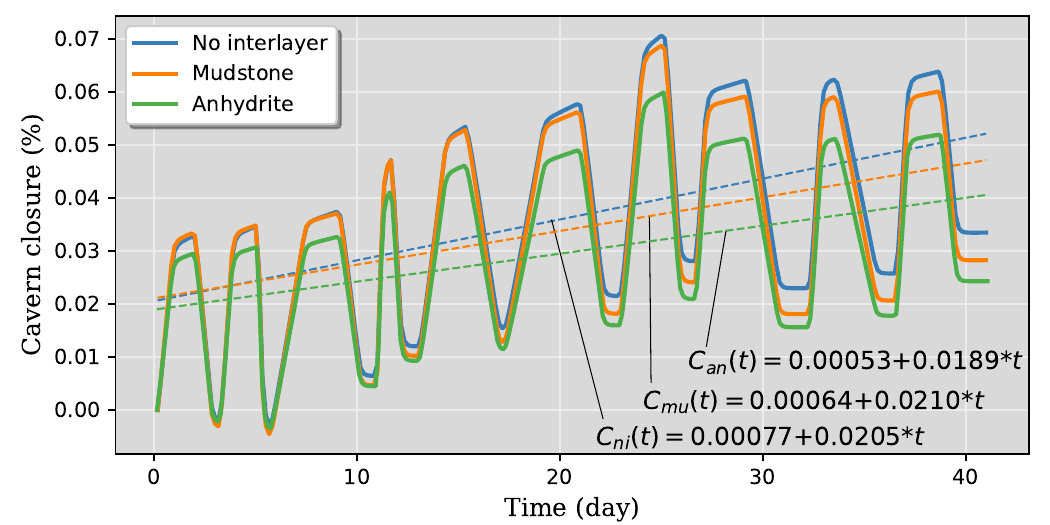}
	\caption{Cavern closure obtained with no interlayer, mudstone interlayer and anhydrite interlayer. Fitted equations $C_{an}(t)$, $C_{mu}(t)$ and $C_{ni}(t)$ consider time in days.}
    \label{fig:fig_17}
\end{figure}

\subsection{Impact of interlayers (heterogeneity)}
Although homogeneous distributions of halite are common in salt domes, the lithology of bedded salt formations can become considerably more complex, with the presence of different types of insoluble interlayers such as anhydrite, gypsum, mudstone, etc \cite{yin2020stability,reedlunn2018enhancements,zhao2023creep}. In this section, we study the impact of the presence of anhydrite and mudstone interlayers on the volume loss of a cavern operating under the pressure schedule S3 (see Fig. \ref{fig:fig_5}). For comparison, this problem is solved with three different property distributions: (i) a homogeneous distribution of halite (Salt-A in Table \ref{tab:mat_params}) in the entire domain; and halite (Salt-A) with an interlayer of (ii) mudstone or (iii) anhydrite crossing the Cavern B of Fig. \ref{fig:fig_4}. Both anhydrite and mudstone are assumed to be purely elastic with the properties shown in Table \ref{tab:mat_params}. Figure \ref{fig:fig_17} shows the cavern closure obtained for the three cases. Because mudstone is softer (lower Young's modulus) than the halite, the amplitude of the cavern closure oscillations with the mudstone interlayer case is slightly bigger than that without interlayer. The fact that the mudstone layer is considered not to undergo creep contributes to reduce cavern closure when compared to the purely halite distribution. Conversely, the amplitude of oscillations for the anydrite interlayer case is smaller than the other cases due to its higher stiffness. The presence of an anhydrite layer is also shown to inhibit cavern closure even more than the mudstone case.

The difference between the three curves in Fig. \ref{fig:fig_17} after 40 days of operation is not significant. In order to predict the difference between these solutions in the long term, linear equations are fitted using least squares, resulting in equations $C_{an}(t)$, $C_{mu}(t)$ and $C_{ni}(t)$ for the cases with anhydrite interlayer, mudstone interlayer and with no interlayer, respectively. Assuming the loading conditions remain approximately the same as in pressure schedule S3 for the next 100 years, these three equations predict volume losses of 19.2\%, 23.2\% and 28.1\% for anhydrite, mudstone and no interlayer, respectively. In other words, the differences between cavern closures obtained by considering or not the interlayers after 100 years of operation are approximately $5\%$ for mudstone interlayer and $9\%$ for the anhydrite interlayer when compared to the no interlayer case.

\subsection{System of caverns}
To investigate the mechanical stability of system of caverns, we consider the geometries illustrated in Fig. \ref{fig:fig_6} and analyze the cavern closures and FOS for each configuration. The pressure schedules applied to the caverns are illustrated in Fig. \ref{fig:fig_7}, which shows that Cavern 2 is kept at a constant pressure of 9 MPa for 192 hours, whereas a cyclic loading is imposed on Cavern 1 during the entire simulation. After 192 hours, Cavern 2 is also subjected to the same cyclic loading in synchronicity with Cavern 1. In this manner, it is possible to study how the operation of one cavern affects the operation of the other. 

Figure \ref{fig:fig_17} shows the cavern closure results of both caverns for the 4 cavern configurations. For the case where $L=0.5R$, it is possible to notice some oscillations in Cavern 2 due to the operations in Cavern 1 during the inactive period of the former. These oscillations are not observed for the other cases and are attributed to the excessive proximity between the two caverns. Another interesting fact is that the cavern closure rate increases after Cavern 2 starts to operate, which can be observed in both $L=0.5R$ and $L=1R$ but not for larger pillar widths. Moreover, although mutual interaction does not seem to take place for $L=2R$, it is possible to notice that after 20 days the cavern closure for this case is larger than for $L=10R$. This is related to the narrower pillar width available to sustain the overburden, resulting in higher stresses and thus higher cavern closure. It is important to point out that the results shown in Fig. \ref{fig:fig_17} are in agreement with \cite{wang2015allowable}, which establishes that the minimum allowable pillar width should be between 2 to 2.5 times the cavern diameter.

\begin{figure}[!h]
	\centering
	\includegraphics[scale=0.54]{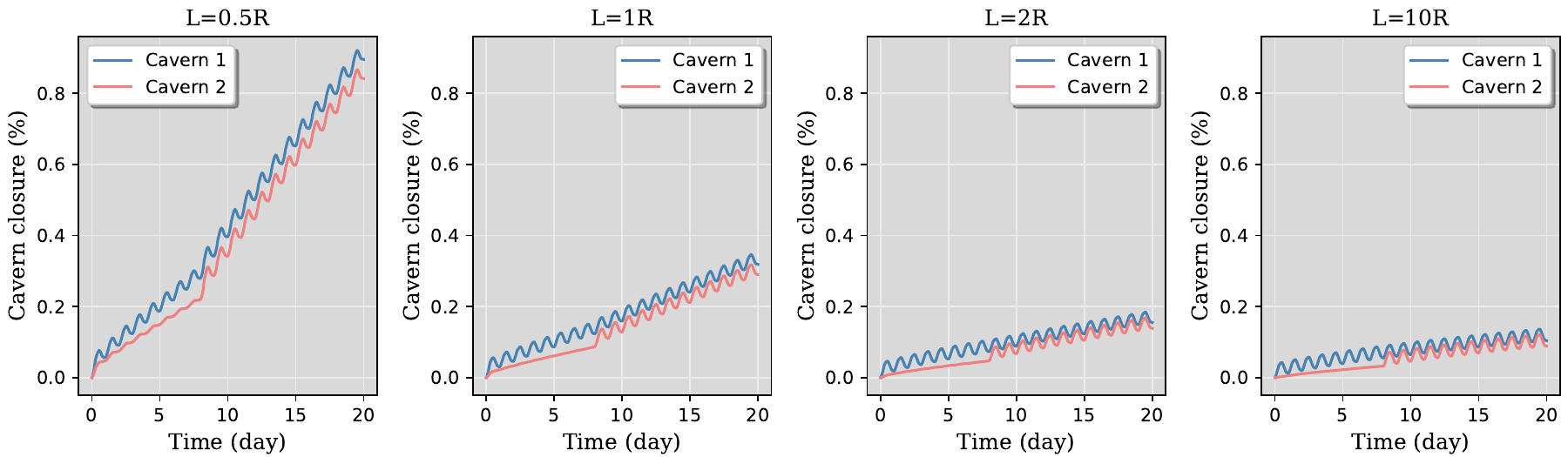}
	\caption{Cavern closure for system of caverns considering different distances between neighbor caverns.}
    \label{fig:fig_17}
\end{figure}

\begin{figure}[!t]
	\centering
	\includegraphics[scale=0.4]{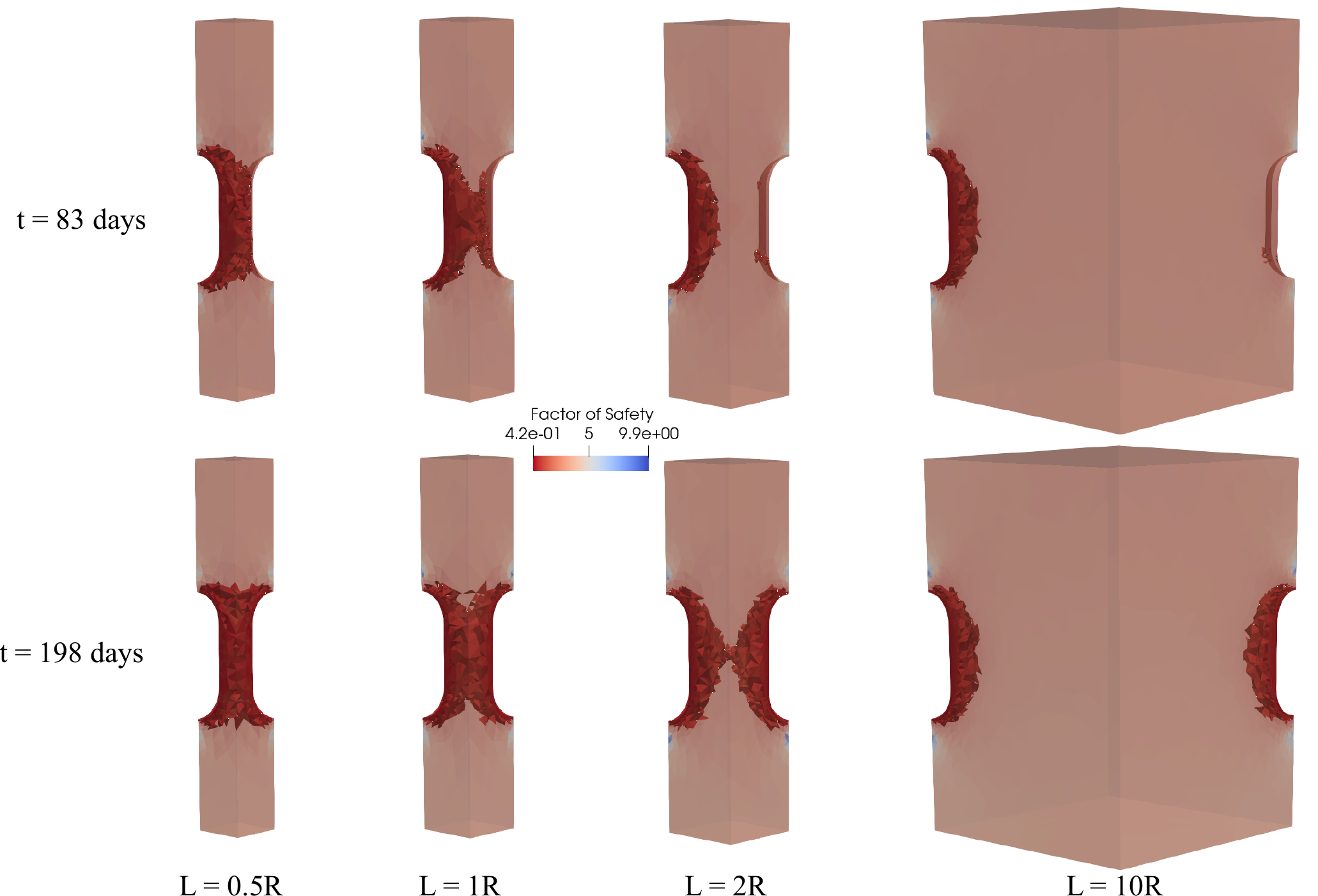}
	\caption{Factor of safety distribution highlighting in red the elements with high probability of presenting tertiary creep (FOS$\leq1.0$).}
    \label{fig:fig_19}
\end{figure}

The stress fields obtained are used to compute the factor of safety (FOS), as described in subsection \ref{subsec:fos}, for each element of the grid. The results are shown in Fig. \ref{fig:fig_19} for when only Cavern 1 is operating ($t=83$ days) and for when both caverns are in operation ($t=198$ days). Times $t=83$ days and $t=192$ days are indicated in Fig. \ref{fig:fig_7}. The red elements indicate where tertiary creep is most likely to occur (FOS$\leq1.0$). For $t=83$ days, a clear interaction between the two caverns is observed for $L=0.5R$ and $L=1R$, but Cavern 2 is not significantly affected for $L=2R$ and $L=10R$. When Cavern 2 is also in operation, the elements with FOS$\leq1.0$ meet in the middle of the two caverns, indicating mutual interaction for $L=2R$. Although tertiary creep is induced in both caverns during the double cyclic operation period, mutual interaction is not observed for $L=10R$.

\section{Conclusions}
\label{sec:conclusion}
This work presented 3D simulation and sensitivity analyses of salt caverns for energy storage, with particular attention devoted to cyclic loading conditions. A comprehensive constitutive model composed of elastic, viscoelastic, viscoplastic and dislocation creep was proposed. The finite element method was employed to solve the non-linear set of equations on unstructured grids. The formulation allows for different time integration schemes (explicit, Crank-Nicolson and fully-implicit) to be used. The analyses aimed to investigate the influence of model calibration, different deformation mechanisms, the impact of cavern geometry and non-salt interlayers, and the mutual interaction in systems of caverns. The main conclusions are summarized below:

\begin{itemize}
    \item Model calibration should be carefully performed to avoid misleading simulation results. For complex constitutive models, there might be different sets of material parameters that present a reasonably good fit against laboratory experiments but lead to very different performance analyses at the salt-cavern scale.
    \item Viscoplastic deformation, describing transient creep, was found crucial in reproducing the laboratory experiments on rock salt specimens (cm-scale). However, for the cases studied in this work, the results suggest that viscoplasticity does not play a crucial role in cavern-scale deformation simulations under cyclic loading. Nevertheless, the stress path is significantly affected by viscoplasticity. It must be emphasized that this remark is valid for the specific viscoplastic model employed in this work. Other models might lead to different conclusions depending on the shape of yield surface, hardening rule, etc.
    \item After a pressure drawdown (or build-up) stage, the viscoelastic deformations (i.e. reverse creep) is important for a relatively short period of time, i.e., approximately 10 hours. After this period, the effects of reverse creep vanish and no cumulative impacts are observed in the long-term.
    \item Irregular cavern shapes do not necessarily induce more volume loss, but they can increase the likelihood of tertiary creep to occur. Interestingly, sharp angles are shown to favor a faster stress redistribution, thus abbreviating the period of tertiary creep. In other words, cavern shapes with sharp angles induce more tertiary creep, but for a shorter period.
    \item Although the presence of interlayers has an impact on the simulation results, the influence is not as significant as the model calibration. Moreover, it was observed that stiffer interlayers decrease the speed of cavern volumetric closure.
    \item By analyzing the factor of safety, mutual interaction between neighboring caverns can be observed for pillar width less than 2 times the cavern radius. Therefore, larger distances should be considered when designing safe systems of caverns for large-scale energy storage.
\end{itemize}

\section*{CRediT authorship contribution statement}
 
\textbf{H.T.H.}: Conceptualization, Methodology, Software, Validation, Formal analysis, Visualization, Writing -- Original Draft.
\textbf{H.H.}: Conceptualization, Methodology, Writing -- Review \& Editing.

\section*{Declaration of competing interest}
The authors declare that they have no known competing financial interests or personal relationships that could have appeared to influence the work reported in this paper.

\section*{Data availability}
Digital data sets of the results and input data are available upon request.

\section*{Acknowledgments}
This research was partly supported by Shell Global Solutions International B.V. within the project ``SafeInCave''. The developments of this study have been made in an open-source ``SafeInCave'' simulator, which can be shared with public on request. The authors also acknowledge members of the ADMIRE and DARSim research groups at TU Delft for the fruitful discussions during the development of this work. Additionally, the authors acknowledge the fruitful discussions with the staff of Shell Global Solutions International B.V., namely, Dr. Maartje Houben, Dr. Kevin Bisdom, Dr. Thomas Fournier, and Dr. Karin de Borst. Also, fruitful discussions with Prof.dr.ir. Lambertus Sluys of TU Delft are acknowledged.

\appendix

\bibliographystyle{unsrt}
\bibliography{main}

\end{document}